\documentclass[11pt]{article}

\usepackage{epsfig}
\usepackage{amsmath}
\usepackage{enumerate}
\usepackage{amssymb}
\usepackage{textcomp} 

\newcommand{\braket}[1]{\langle #1 \rangle}
\newcommand{\Tprod}[1]{{\mathrm T}\lbrack #1 \rbrack}
\newcommand{\ket}[1]{| #1 \rangle}
\newcommand{\permille}{\textperthousand}
\def\re{\mathrm{Re}}
\def\im{\mathrm{Im}}

\allowdisplaybreaks
\begin{document}

\begin{titlepage}

  \begin{flushright}
    \normalsize PITHA~08/13\\
    \normalsize SFB/CPP-08-34\\
    \normalsize June 30, 2008\\    
  \end{flushright}

\vskip1.5cm

\begin{center}
  \Large\bf\boldmath
    Dominant NNLO corrections to four-fermion production\\
    near the $W$-pair production threshold
  \unboldmath
\end{center}

\vspace*{0.8cm}

\begin{center}
  {\sc S. Actis},
  {\sc M. Beneke}, 
  {\sc P. Falgari},
  {\sc C. Schwinn}\\[5mm]
  
  {\small\it Institut f{\"u}r Theoretische Physik~E, RWTH Aachen University,}\\
  {\small\it D--52056 Aachen, Germany}\\[0.1cm]        
\end{center}

\vspace*{0.8cm}

\begin{abstract}
  \noindent
  We calculate the parametrically dominant next-to-next-to-leading
  order corrections to four-fermion production $e^- e^+\to \mu^-
  \overline{\nu}_\mu u \overline{d}+X$ at centre-of-mass energies near
  the $W$-pair production threshold employing the method of
  unstable-particle effective theory. In total the correction is
  small, leading to a shift of 3$\,$MeV in the $W$-mass measurement.
  We also discuss the implementation of realistic cuts and provide a
  result for the interference of single-Coulomb and soft radiative
  corrections that can easily be extended to include an arbitrary
  number of Coulomb photons.
\end{abstract}

\vfil

\end{titlepage}

\newpage

\section{Introduction}
$W$-pair production at $e^- e^+$ colliders is a key process for
measuring the $W$-boson mass and testing the non-abelian structure of
the Standard Model (SM), because of its sensitivity to triple
non-abelian gauge couplings. The total cross section has been measured
at LEP2 in a kinematical region spanning from the $W$-pair production
threshold to a centre-of-mass energy of $207$ GeV~\cite{LEP2} with an
accuracy of $\sim \! 1\%$ at the highest energies. The $W$ mass has
been determined with an error of $\sim \! 40$ MeV reconstructing the
$W$ bosons from their decay products; bounds on possible anomalous
couplings are less stringent, and deviations from the SM predictions
have been constrained at the per-cent level. A detailed analysis of
the $W$-pair production process will be possible at the planned
International Linear Collider, where the total cross section could be
measured at the per-mille level~\cite{AguilarSaavedra:2001rg}. The
precision on the $W$-mass determination has been estimated to be $\sim
\! 10$ MeV by a direct reconstruction of the $W$-decay
products~\cite{MoenTonazzo}, and $\sim \! 6$ MeV by a dedicated
threshold scan~\cite{Wilson}. Moreover, the possibility to scan the
high-energy region would allow to measure more precisely the size of
the triple non-abelian gauge couplings.

The estimates on the $W$-mass determination rely on statistics and the 
performance of the future linear collider, and they assume that the cross 
section can be predicted by theory to sufficient accuracy in order to 
convert its measurement into one for the $W$ mass. For this 
reason, radiative corrections to on-shell $W$-pair production have been 
thoroughly investigated in the past and are known at next-to-leading 
order (NLO) since the beginning of the 1980's~\cite{Lemoine:1979pm}. 
However, the $W$ bosons being unstable, a precise theoretical prediction  
has to be formulated for a final state of stable or sufficiently long-lived 
particles, represented by the fermion pairs produced by $W$ decay, rather 
than for on-shell $W$ bosons. NLO predictions for four-fermion production 
far from the $W$-pair threshold region are available since some time  in the 
double-pole approximation~\cite{Beenakker:1998gr,Denner:1999kn,Denner:2000bj} 
or with further simplifications~\cite{Jadach:2000kw,Jadach:2001uu}. Recently, 
a full NLO computation of four-fermion production has been performed in the 
complex-mass scheme~\cite{Denner:2005es,Denner:2005fg} without any kinematical 
approximation; moreover, a compact analytic result around the threshold region 
has been obtained in~\cite{Beneke:2007zg} using effective field theory (EFT) 
methods~\cite{Beneke:2003xh,Beneke:2004km,Chapovsky:2001zt}.

In particular, the analysis performed in~\cite{Beneke:2007zg}
led to the following conclusions: 1) a resummation of next-to-leading
collinear logarithms from initial-state radiation is mandatory to reduce
the error on the $W$ mass from the threshold scan 
below $\sim$ $30$ MeV; 2) the NLO partonic
cross-section calculation in the EFT approach implies a residual error
of $\sim$ $10-15$ MeV. Although a large component of the 
uncertainty at point 2) can be removed using the 
full NLO four-fermion calculation~\cite{Denner:2005es,Denner:2005fg}, 
the computation of the dominant higher order corrections,
whose contribution has been estimated to be roughly $\sim$ $5$ MeV 
in~\cite{Beneke:2007zg}, is necessary to secure 
the $6$ MeV accuracy goal~\cite{Wilson}.

In this paper we employ EFT techniques to calculate analytically the
(parametrically) dominant next-to-next-to-leading order (NNLO)
corrections to the inclusive four-fermion production process
$e^-e^+\to \mu^-\overline{\nu}_\mu u\overline{d}+X$, where $X$ stands
for an arbitrary flavour-singlet state, with a three-fold goal: to
improve the EFT NLO calculation~\cite{Beneke:2007zg}; to derive a
result which can be added on top of the full NLO
prediction~\cite{Denner:2005es,Denner:2005fg}; to reduce the
uncertainty on the $W$-mass measurement below the required $5\,$MeV
level by including a new set of higher order corrections.

The organisation of the paper is as follows. In section~\ref{sec:outline} we 
outline the structure of our computation, review the essential 
features of the EFT method and identify the set of parametrically 
leading NNLO corrections. In section~\ref{sec:eval} we describe in
detail the calculation of the various contributions.
In section~\ref{sec:nums} we show the numerical impact of our result 
on the inclusive cross section and in section~\ref{sec:cuts}
we discuss the effect of realistic cuts adopting those applied 
at LEP2 as a template. As a by-product we explain how invariant-mass 
cuts can be included in the EFT approach.
Finally, section~\ref{sec:conc} contains our conclusions and 
two appendices collect results related to the renormalisation of 
the Coulomb potential and the electromagnetic coupling, and the
conversion from the fixed-width to the complex-mass scheme.
\section{Outline of the computation}
\label{sec:outline}
We consider the inclusive four-fermion production process
  \begin{equation}
  \label{cor:proc}
    e^-(p_1)\,e^+(p_2)\,\to\, 
    \mu^-\,\overline{\nu}_\mu\, 
    u\,\overline{d}\,+\,X,
  \end{equation}
where $X$ denotes an arbitrary flavour-singlet state (nothing, photons, 
gluons, \ldots), in the kinematical regime close to the $W$-pair production 
threshold, $s\!\equiv\!(p_1+p_2)^2\!\sim \!4\,M_W^2$. Here the total cross 
section is dominated by the production of two resonant non-relativistic 
$W$ bosons with virtuality of order $ k^2\,-\,M_W^2 \sim  M_W^2\, v^2 \sim 
M_W\,\Gamma_W \ll M_W^2$, where $v$ is the non-relativistic velocity and 
$M_W$ and $\Gamma_W$ are the $W$ pole mass and decay width. We recall that 
the relation between the pole mass and the value given by experimental 
collaborations $\hat{M}_W=(80.403\pm 0.029)$ GeV~\cite{Yao:2006px}  is 
given by $\hat{M}_W \!-\! M_W\!=\!\Gamma_W^2\slash (2M_W) + {\cal O} 
(\alpha_{ew}^3)$~\cite{Sirlin:1991fd}, where $\alpha_{ew}\equiv \alpha \slash s^2_w$, 
$\alpha$ is the fine-structure constant, $s^2_w\equiv \sin^2\theta_w$ and 
$\theta_w$ denotes the weak-mixing angle.

In the framework of unstable-particle effective field theory 
(EFT)~\cite{Beneke:2003xh,Beneke:2004km,Chapovsky:2001zt} the total cross 
section is obtained through a re-organised loop expansion and a kinematical 
expansion in the small parameters
  \begin{equation}
    \alpha_{ew},\qquad 
    \frac{s\,-\,4\, M_W^2}{4\, M_W^2}\sim v^2,\qquad 
    \frac{\Gamma_W}{M_W}\sim \alpha_{ew}.
  \end{equation}
All three parameters are of the same order, and in the following we 
will denote them as $\delta$ for power-counting purposes. Obviously, the
EFT expansion is not equivalent to a standard loop expansion; 
in particular, being 
tailored to a specific kinematical region, it allows for a 
straightforward evaluation 
of leading higher order corrections. In order to appreciate 
this point, let us briefly 
review how the EFT exploits the hierarchy of scales and deals 
with the threshold region.
The key observation consists in recognising four momentum 
scalings in the centre-of-mass 
frame, in the spirit of the method of regions~\cite{Beneke:1997zp},
  \begin{equation}
  \label{scale}
  \begin{aligned}
    \text{hard }      &: \quad k_0\sim |\vec{k}|\sim M_W,          \qquad\,\,
    \text{potential }  : \quad k_0\sim M_W\, \delta,               \quad\,
                               |\vec{k}|\sim M_W\, \sqrt{\delta},  \\
    \text{soft }      &: \quad k_0\sim |\vec{k}|\sim  M_W\, \delta,\qquad
    \text{collinear }  : \quad k_0 \sim M_W,                       \qquad 
                               k^2\sim M_W^2\delta,
  \end{aligned}
  \end{equation}
where $k$ is an arbitrary loop-integration momentum.
Starting from NNLO diagrams another mode has to be
included in addition to the ones given
in~\eqref{scale} that will be called\footnote{This is 
analogous to the `soft' mode in the  NRQCD literature 
whose `ultrasoft' mode corresponds to the
`soft' mode in our conventions.} `semi-soft':
\begin{equation}
\label{eq:semi-soft}
  \text{semi-soft} \quad:\quad k_0\sim |\vec k|\sim M_W\sqrt \delta.
\end{equation}
Hard modes are integrated out,
and their contribution is encoded in appropriate matching
coefficients. The remaining
dynamical modes appear in genuine EFT loop computations and their different
scaling properties lead to a non-standard power counting in the 
small expansion parameter $\delta$ involving half-integer powers.

  \begin{figure}[ht]
  \begin{center}
  \includegraphics[width=0.9\textwidth]{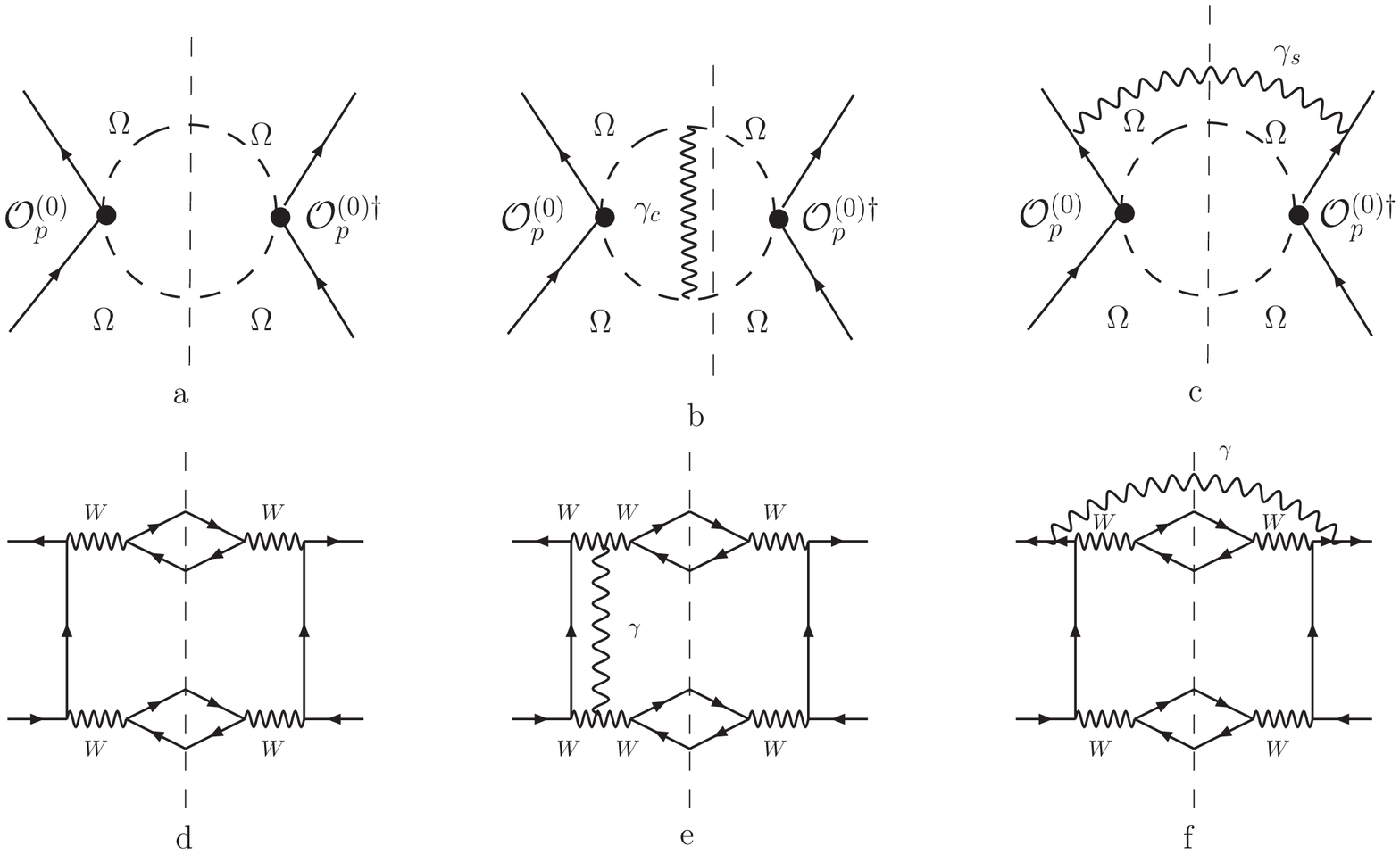}
  \caption{Sample diagrams in the EFT (first line) and in
           the full SM (second line). External fermionic lines are electrons
           and positrons; internal fermionic lines appearing in the diagrams
           in the second line represent the $\mu^-\,\overline{\nu}_\mu$ and
           $u\, \overline{d}$ doublets.}
  \label{fig:LOcount}
  \end{center}
  \end{figure}
The total cross section of~\eqref{cor:proc} is related through unitarity  
to the appropriate cuts of the $e^-\, e^+$ forward-scattering amplitude.
Figure~\ref{fig:LOcount}d shows a sample diagram contributing to the cut
forward-scattering amplitude in the full Standard Model (SM). 
In the EFT language, the leading-order (LO) EFT contribution to the 
forward-scattering amplitude for the helicity configuration 
$e^-_L\, e^+_R$, the only non-vanishing one at LO in the
non-relativistic expansion, reads
  \begin{equation}
  \label{LOamp}
    i {\cal A}^{(0)}_{LR}\,=\,\int d^4 x\,
    \braket{e^-_L e^+_R |\,\Tprod{\,i\, {\cal O}_p^{(0)\dagger }(0)\,
    i\,{\cal O}_p^{(0)}(x)}\,|e^-_L e^+_R }.
  \end{equation}
The corresponding cut diagram is illustrated in figure~\ref{fig:LOcount}a.
Here the operators ${\cal O}_p^{(0)}$ and ${\cal O}_p^{(0)\dag}$ account at LO
for the production and destruction of a pair of non-relativistic $W$
bosons. Their explicit expressions are given by
  \begin{equation}
  \label{cor:ver}
    {\cal O}_p^{(0)}\, =\, \frac{\pi\,\alpha_{ew}}{M_W^2}\,
    \left(\,\bar{e}_{c_2,L}\,  \gamma^{[i} n^{j]}\,  e_{c_1,L}\, \right)\,
    \left(\,\Omega_-^{\dagger i}\, \Omega_+^{\dagger j}\,\right),
  \end{equation}
where we have introduced the short-hand notation $a^{[i} b^{j]}\equiv a^i b^j 
+ a^j b^i$. Here $\vec{n}$ represents the unit vector for the direction of the 
incoming electron three-momentum $\vec{p}_1$, $\gamma^i$, with
$i=1,2,3$, are the usual Dirac matrices, the subscripts on the 
electron fields denote the two different direction labels of the
collinear fields and $\Omega_\pm^i$, with $i=1,2,3$, are 
non-relativistic spin-1 destruction fields for particles with electric charge 
$\pm 1$, whose propagators are
  \begin{equation}
  \label{cor:delta}
    \frac{i\, \delta^{ij}}{k^0\,-\,\frac{|\vec{k}|^2}{2 M_W}\,+\, 
    \frac{i\, \Gamma_W}{2}}.
  \end{equation}
Loop corrections to the unstable-particle EFT scattering amplitude are then
computed using the vertex~\eqref{cor:ver}, 
the propagator~\eqref{cor:delta} and interactions with potential
photons, related to the Coulomb force between the slowly-moving $W$ 
bosons, (semi-)~soft, and collinear photons.

For definiteness, let us consider selected radiative corrections to 
the LO EFT cut diagram from figure~\ref{fig:LOcount}a, and show 
how the non-standard power counting arises. From the scaling 
properties~\eqref{scale},
it is clear that the propagators~\eqref{cor:delta} of the potential 
fields $\Omega$ 
scale as $1\slash\delta$ and the integration measure $d^4\,k$ as 
$\delta^{5\slash 2}$; 
since each vertex~\eqref{cor:ver} contains the coupling constant 
$\alpha_{ew}$, we 
can associate a power-counting factor 
$\alpha_{ew}^2\,\delta^{1\slash 2}$ to the LO diagram.

We now turn to virtual Coulomb (potential) corrections, shown in 
figure~\ref{fig:LOcount}b, where we denote the loop momentum of 
the Coulomb photon by $k_c$. The corresponding SM diagram is depicted 
in figure~\ref{fig:LOcount}e. The photon propagator scales as 
$1\slash \vec{k}_{c}^{\,2}\! \sim \! 1\slash \delta$ and the integration
measure $d^4 k_c$ as $\delta^{5\slash 2}$; including the overall 
factor $\alpha \sim \delta$ due to the photon interaction vertices and
two additional $W$ propagators, we obtain a power-counting 
factor $\alpha_{ew}^2\delta$. The situation 
is quite different for the soft real-photon corrections illustrated in 
figure~\ref{fig:LOcount}c (see figure~\ref{fig:LOcount}f for the 
Standard Model counterpart). The photon propagator, 
parametrised by the loop momentum $k_s$, scales as 
$1\slash {k}_{s}^2 \sim 1\slash \delta^2$, and the integration 
measure $d^4k_s$ as $\delta^{4}$; each collinear internal fermion 
line scales as $1\slash (k_s\cdot p_i)\sim 
1\slash \delta$, with $i=1,2$, where $p_1$ and $p_2$ are the 
incoming momenta of the external electron and positron; 
the two photon vertices introduce an overall factor $\alpha \sim \delta$; 
the power-counting factor associated to the soft-photon 
diagram is $\alpha_{ew}^2\,\delta^{3\slash 2}$. 

The standard loop expansion of figure~\ref{fig:LOcount}e and
figure~\ref{fig:LOcount}f treats virtual Coulomb corrections and real 
soft-photon contributions as genuine NLO
effects. The re-organised EFT expansion, instead, shows 
that Coulomb-photon corrections at the $W$-pair production threshold 
are suppressed with respect to the LO diagram by a factor 
$\delta^{1\slash  2}$, and can be classified as dominant effective 
N$^{1\slash 2}$LO$^{\text {EFT}}$ terms in the small 
expansion-parameter $\delta$; soft-photon corrections are suppressed 
by one power of $\delta$, and can be treated as sub-dominant 
NLO$^{\text {EFT}}$ effects.

  \begin{figure}[t]
  \begin{center}
  \includegraphics[width=0.9\textwidth]{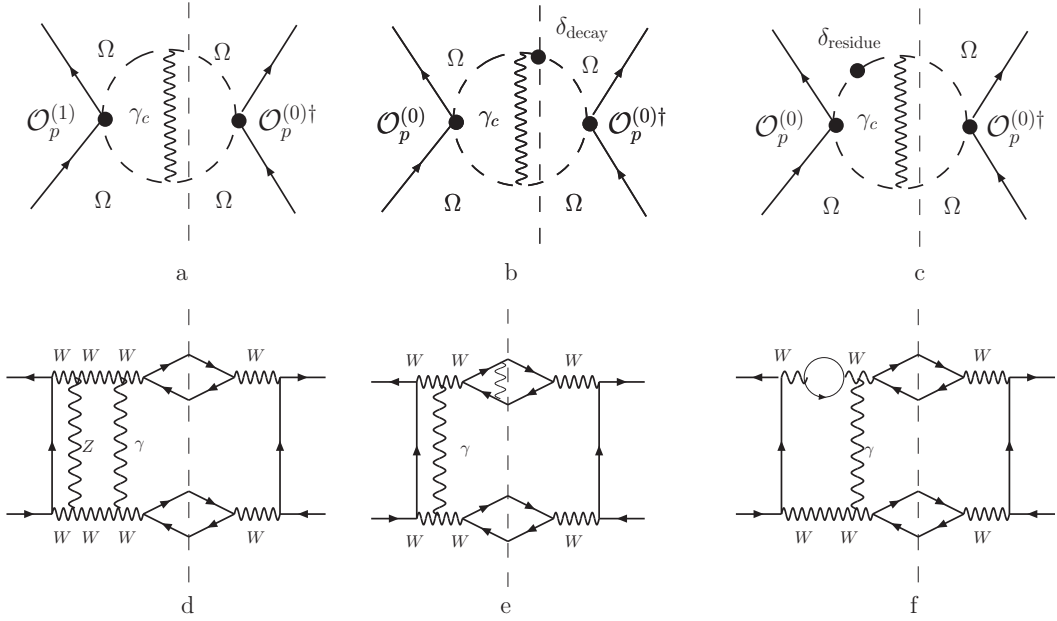}
  \caption{Sample N$^{3\slash 2}$LO$^{\text {EFT}}$ diagrams in the
    EFT (first line) with mixed hard/Coulomb corrections and
    corresponding NNLO diagrams in the full SM (second line). Note the
    insertion of a NLO production operator in diagram a. The
    same conventions of figure~\ref{fig:LOcount} are adopted.}
  \label{fig:NLOcount}
  \end{center}
  \end{figure}
\begin{figure}[ht]
  \begin{center}
  \includegraphics[width=0.9\textwidth]{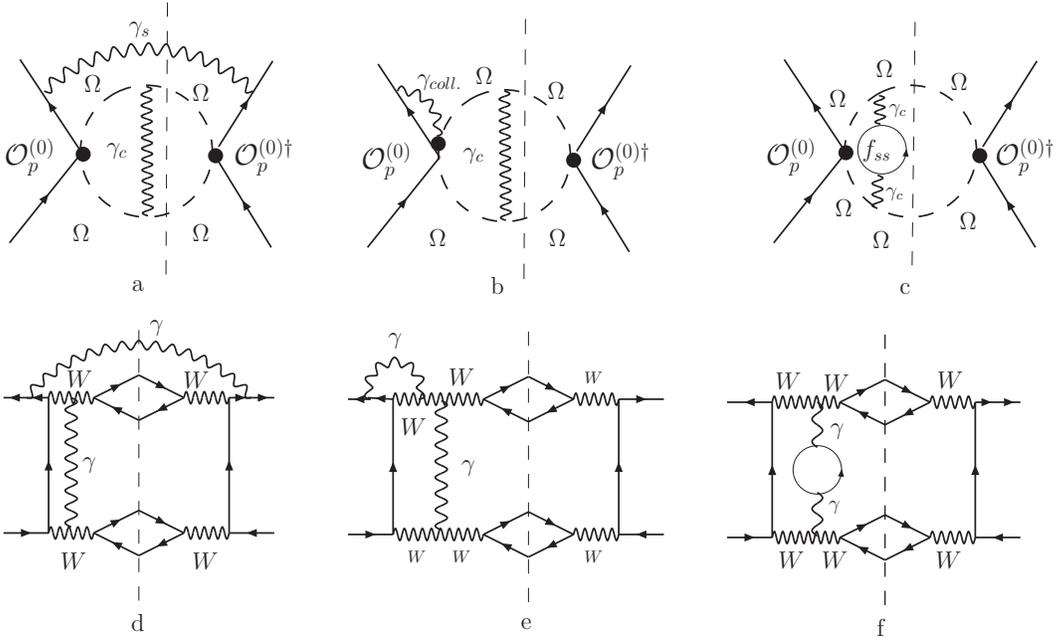}
  \caption{Sample N$^{3\slash 2}$LO$^{\text {EFT}}$ diagrams with single-Coulomb exchange and radiative corrections in the
    EFT (first line) and
    corresponding NNLO diagrams in the full SM (second line).
The
    same conventions of figure~\ref{fig:LOcount} are adopted.}
  \label{fig:NLOcount2}
  \end{center}
  \end{figure}

  The aim of this paper is to calculate corrections to the
  four-fermion production cross section which are not included in the
  SM NLO result, but might be relevant to the $W$ mass analysis at the
  $5\,$MeV level. It is therefore natural to turn our attention to
  that subset of the SM NNLO diagrams
  which corresponds to EFT terms containing an extra factor
  $\delta^{3\slash 2}$ with respect to the LO result, rather than
  $\delta^2$.  Corrections of the same order in the EFT power counting
  that are already included in the SM LO or NLO diagrams (non-resonant
  contributions encoded in four-electron operators in the EFT; higher
  dimensional production operators; etc.) are not considered in the
  present paper.  Those N$^{3\slash 2}$LO$^{\text {EFT}}$ radiative
  corrections that correspond to SM NNLO diagrams can be readily
  organised in several classes:
\begin{description}
\item[Mixed hard/Coulomb corrections:] This class is given by diagrams with a  single-Coulomb photon and one insertion of a hard NLO correction to the
            \begin{itemize}
            \item {\it production stage}: Here the LO
              operator~\eqref{cor:ver} is replaced by the NLO
              expression (figure~\ref{fig:NLOcount}a). 
              A representative diagram of the full-SM counterpart is
              shown in figure~\ref{fig:NLOcount}d;
              this correction is computed in section~\ref{sec:hard}.  
          \item {\it decay stage:} 
               A sample diagram in the
              standard loop expansion is shown in figure~\ref{fig:NLOcount}e.
The implementation of this correction in the EFT discussed in
section~\ref{sec:decay}
is denoted by the black dot labelled $\delta_{\text{decay}}$ 
in figure~\ref{fig:NLOcount}b. 
\item{\it propagation stage:}     A sample diagram in the
              standard loop expansion is shown in figure~\ref{fig:NLOcount}f.
The implementation of this correction in the EFT discussed in
section~\ref{sec:residue}
is denoted by the black dot labelled $\delta_{\text{residue}}$ 
in figure~\ref{fig:NLOcount}b. 
            \end{itemize}
\item [Interference of Coulomb and radiative corrections:]
There are two contributions in this class:
  \begin{itemize}
  \item {\it Single-Coulomb exchange and soft photons.} An EFT diagram in
    this class is shown in figure~\ref{fig:NLOcount2}a. A representative
    diagram of the full-SM counterpart is shown in
    figure~\ref{fig:NLOcount2}d.  This correction is computed in
    section~\ref{sec:soft}.
 
\item {\it Single-Coulomb exchange and collinear photons.} 
  These corrections (see
  figure~\ref{fig:NLOcount2}b for a representative diagram in the EFT,
  and figure~\ref{fig:NLOcount2}e for a counterpart in the full SM)
  vanish if the electron mass is set to zero. If a finite electron
  mass is used as infrared~(IR) regulator there are further contributions from 
  soft-collinear and hard-collinear modes that are computed in
  sections~\ref{sec:soft} and~\ref{sec:hard}, respectively.
\end{itemize}
\item[NLO corrections to the Coulomb potential:] The relevant diagram
  is given by a semi-soft fermion bubble insertion into the Coulomb 
  photon (see figure~\ref{fig:NLOcount2}c for the EFT diagram and 
  figure~\ref{fig:NLOcount2}f for the counterpart in the full SM). 
  This correction is computed in section~\ref{sec:bubble} and 
  appendix~\ref{sec:hard-coulomb}.
\end{description}
This concludes our survey of the dominant NNLO corrections. In the
next section we will compute the different contributions identified
above.  At the same order in the EFT power counting as the above
contributions we also encounter triple-Coulomb exchange that is a
NNNLO correction in the standard loop expansion. This effect can be
straightforwardly included using EFT methods but turns out to be
numerically negligible.

\section{Evaluation of the dominant NNLO corrections}
\label{sec:eval}
In section~\ref{sec:outline}  the
dominant NNLO corrections to four-fermion production near the $W$-pair
production threshold have been identified.  In this section we compute all the relevant
contributions.
Subsection~\ref{sec:c3} reviews the known result for the all-order Coulomb Green function that will enter subsequent calculations 
and extracts the triple-Coulomb exchange correction.
Interference of
single-Coulomb exchange with corrections from soft and
soft-collinear photons are computed in subsection~\ref{sec:soft}.
  Subsection~\ref{sec:hard} contains the results for interference of
single-Coulomb exchange with hard corrections to the production  stage
and with hard-collinear photon exchange.
 These computations are carried out in a form that
allows in principle to include all-order Coulomb exchange as well. 
This subset of NNLO corrections is combined to a finite partial result
in subsection~\ref{sec:total} and  large logarithms of the electron mass are
absorbed in electron structure functions in subsection~\ref{sec:isr}. 
Radiative corrections to the single-Coulomb exchange potential itself
are discussed in subsection~\ref{sec:bubble}. Interference effects of
single-Coulomb exchange with corrections to the decay and propagation
stages are subject of subsections~\ref{sec:decay} and~\ref{sec:residue}.

\subsection{Coulomb Green function and triple-Coulomb exchange}
\label{sec:c3}
\begin{figure}[t]
  \begin{center}
  \includegraphics[width=0.9\textwidth]{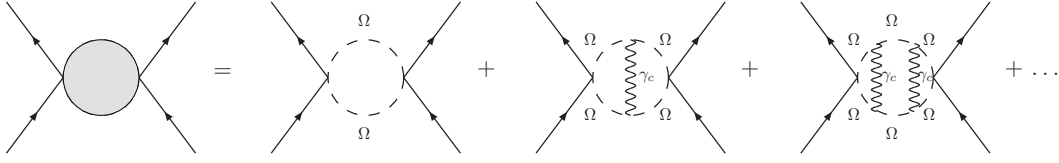}
  \caption{Pure Coulomb corrections to the forward-scattering amplitude.}
  \label{fig:Coulomb}
  \end{center}
  \end{figure}

It is clear that Coulomb corrections will play a privileged role in our
computation. In particular, in the following, we will derive a result which 
can be applied to diagrams involving an arbitrary number of Coulomb photons.
Therefore, in order to illustrate our formalism, we review here the computation of pure Coulomb
corrections to the four-fermion production process~\eqref{cor:proc},
given by the exchange of potential photons with energy $k_0\,\sim\, M_W\delta$ 
and three-momentum $|\vec{k}|\,\sim\, M_W\sqrt{\delta} $. Note that these corrections 
correspond to insertions of non-local four-boson interactions in the effective 
Lagrangian of potential non-relativistic QED~\cite{Pineda:1998kn}. 
They can be summed to all orders in perturbation theory, as shown in 
figure~\ref{fig:Coulomb}, in terms of the zero-distance Coulomb Green
function, the Green function $G^{(0)}_{\rm C}(\vec{r}_1,\vec{r}_2;E)$
of the Schr{\"o}dinger operator $-\vec{\nabla}^2/M_W-\alpha/r$ evaluated
at $\vec{r}_1=\vec{r}_2=0$. 

Using the representation of the Green function 
given in~\cite{WichWoo} and defining $E\equiv \sqrt{s}-2M_W$ and
${\cal E}_W\equiv E+i\Gamma_W$, we can  
represent the Coulomb-corrected $e^- e^+$ forward-scattering amplitude 
corresponding to the diagrams in figure~\ref{fig:Coulomb} as
  \begin{equation}
  \label{ampC}
    {\cal A}_{LR}^{\rm C}= 
    \frac{\pi^2 \alpha_{ew}^2}{M_W^4}\,
    \langle p_2 - |n^{[i}\gamma^{j]} | p_1 - \rangle
    \langle p_1 - |n^{[i}\gamma^{j]} | p_2 - \rangle\,
    \, G_{\rm C}^{(0)}(0,0;{\cal E}_W),
  \end{equation}
where we have adopted the canonical helicity notation 
$\ket{p\pm}\equiv\frac{1\pm\gamma^5}{2} u(p)$. The
$\overline{\text{MS}}$-subtracted representation for the zero-distance Coulomb
Green function with a complex argument ${\cal E}_W$, 
including all-order photon 
exchange, reads as follows~\cite{Beneke:1999zr}:
  \begin{equation}
  \label{coulombGF}
    G_{\rm C}^{(0)}(0,0;{\cal E}_W)\!=\! -\frac{M_W^2}{4\pi} \Bigg\{\!
    \sqrt{-\frac{{\cal E}_W}{M_W}} + \alpha \bigg[
    \frac{1}{2}\ln \bigg(\! -\!\frac{4\,M_W {\cal
        E}_W}{\mu^2}\!\bigg)\!-\frac{1}{2} +\gamma_E
    +\psi\bigg(1-\frac{\alpha}{2\sqrt{-{\cal E}_W \slash M_W}}\!\bigg)\!
    \bigg]\!\Bigg\}.
  \end{equation}
Here $\gamma_E$ is the Euler-Mascheroni constant, 
$\mu$ the 't~Hooft unit of mass,
and $\psi$ the Euler psi-function. The total cross section for the 
left-right helicity configuration can be obtained by 
evaluating the spinor product 
$\langle p_2 - |n^{[i}\gamma^{j]} | p_1 - \rangle
\langle p_1 - |n^{[i}\gamma^{j]} | p_2 -
\rangle=16\,(1-\epsilon)\,M_W^2$
and by multiplying the 
imaginary part of the forward-scattering amplitude~\eqref{ampC} by 
the LO branching-fraction 
product $\text{Br}^{(0)}_{W^-\to \mu^- \bar{\nu}_\mu} 
\text{Br}^{(0)}_{W^+ \to u\bar{d}}=
1/27$.
Finally, expanding the psi function appearing in~\eqref{coulombGF} 
in $\alpha$, we get the one- and two-Coulomb photon exchange 
terms~\cite{Fadin:1993kg,Fadin:1995fp} 
  \begin{equation}
  \label{CoulombCS}
    \sigma_{LR}^{\rm C}= - \frac{4\pi\alpha_{ew}^2}{27\,s}\,\mbox{Im}
    \left[\sqrt{-\frac{{\cal E}_W}{M_W}} + 
    \frac{\alpha}{2} \ln \left( - \frac{{\cal E}_W}{M_W}\right)
    - \frac{\alpha^2\pi^2}{12}\sqrt{-\frac{M_W}{{\cal E}_W}}\,
    \right],
  \end{equation}
where the scheme-dependent real constant of~\eqref{coulombGF}
has dropped out. These three terms, shown in figure~\ref{fig:Coulomb},
are already included in the NLO EFT computation.

Triple-Coulomb exchange arises from a NNNLO correction in the standard
loop counting, but contributes at order N$^{3\slash
  2}$LO in the EFT power-counting. 
The result can be obtained straightforwardly by
expanding~\eqref{coulombGF} up to order $\alpha^3$:
  \begin{equation}
  \label{Coulomb3}
   \Delta \sigma^{{\rm C3}}_{LR}= 
   \frac{\pi\alpha_{ew}^2}{27\,s}\,\alpha^3 \zeta(3)\,
    \mbox{Im}\left[- \frac{M_W}{{\cal E}_W} 
    \right].
  \end{equation}
The correction to the helicity-averaged cross section 
$\Delta\sigma^{\rm{C3}} = \Delta\sigma^{\rm C3}_{LR}/4$ 
directly at threshold is $\Delta\sigma^{\rm{C3}}(\sqrt s = 161
\text{GeV}) =0.01$fb while the effect is even smaller away from threshold.

\subsection{Soft and soft-collinear corrections}
\label{sec:soft}
\begin{figure}[ht]
  \begin{center}
  \includegraphics[width=0.6\textwidth]{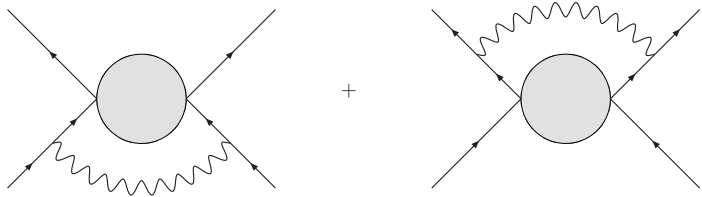}
  \caption{Soft-photon corrections to the all-order Coulomb-corrected 
           forward-scattering amplitude.}
  \label{fig:Soft}
  \end{center}
  \end{figure}

We consider here the radiative corrections obtained including soft-photon 
exchange to the diagrams shown in figure~\ref{fig:Coulomb}. These
${\cal O}(\alpha)$ contributions to the Coulomb-corrected forward-scattering 
amplitude are related to the initial-initial state interference diagrams in the 
effective theory  shown in figure~\ref{fig:Soft}, where the photon energy and
three-momentum scale both as $k_0\sim |\vec{k}|\sim M_W\, \delta$. 
Analogously to the NLO calculation in~\cite{Beneke:2007zg}, diagrams
with a coupling of soft photons to $\Omega$ lines cancel as can be shown 
using a gauge-invariance argument.
The coupling of soft photons to collinear electrons and positrons is 
given by the soft-collinear 
effective-theory (SCET) 
Lagrangian~\cite{Bauer:2000yr,Bauer:2001yt,Beneke:2002ph},
and amounts to the eikonal coupling $\pm i e n^\mu$, where 
$n^\mu$ stands for the 
direction of the four-momentum of the electron or the positron.

The sum of the two diagrams shown in figure~\ref{fig:Soft} can be
written in close analogy with~\eqref{ampC} as 
  \begin{equation}
  \label{TotSoft}
    i {\cal A}_{LR}^{{\rm C}\times {\rm S}}= 
    \frac{16\pi^2\alpha_{ew}^2}{M_W^2}\,
    \left(1-\epsilon\right)\,2\,{\cal I}_{\rm S},
  \end{equation}
where ${\cal I}_S$ denotes the convolution of the zero-distance Green function
and a single soft-photon exchange correction. Including prefactors it reads
  \begin{equation}
  \label{intSoft}
    {\cal I}_S \equiv -32\pi\alpha\,M_W^2\,\tilde \mu^{2\epsilon} \int 
    \frac{d^d k}{(2\pi)^d} \frac{G_{\rm C}^{(0)}(0,0;{\cal E}_W-k^0)}
    {(k^2+i0)(-2 p_1\cdot k +i0)(-2 p_2\cdot k+i0)}.
  \end{equation}
Here the space-time dimension is $d\equiv 4-2\epsilon$, 
$\tilde \mu^2\equiv \mu^2 
e^{\gamma_E}\slash (4\,\pi)$ and $G_{\rm C}^{(0)}$ is given, to all
orders in $\alpha$ but in $d=4$, 
in \eqref{coulombGF}. 

The soft integral \eqref{intSoft} can be evaluated in two steps:
first, we approximate the momenta of the external electron and 
positron introducing 
$p_1\equiv M_W n_+$ and $p_2\equiv M_W n_-$, where $n_+$ and $n_-$ fulfil the 
relations $n_+^2=n_-^2=0$ and $n_+ \cdot n_-=2$; next, we perform the $k^0$ integration 
closing the integration contour in the lower complex $k^0$ half-plane, using Cauchy
theorem and Jordan lemma and picking up the residue of the pole at $k^0= |\vec{k}|-i0$. 
Note that all singularities of $G_{\rm C}^{(0)}$ are located in the upper $k^0$
half plane. The result is
  \begin{equation}
  \label{resSoft}
  {\cal I}_S= -i \,\left(\frac{\alpha}{\pi}\right)\, \frac{\sqrt{\pi}}
  {\epsilon\,\Gamma(1\slash 2 -\epsilon)} (e^{\gamma_E} \mu^2)^\epsilon
  \int_0^\infty\, \!\!\! dk\, \frac{G_{\rm C}^{(0)}(0,0;{\cal E}_W-k)}{k^{1+2\epsilon}}.
\end{equation}

The single pole in the $\epsilon$ plane in the prefactor
of~\eqref{resSoft} is associated with the emission of photons
collinear to the incoming electron or positron, whose mass must be 
neglected in the soft region. The finite electron mass requires 
the introduction of two further regions 
(soft- and hard-collinear)~\cite{Beneke:2007zg}, which convert the 
collinear $1/\epsilon$ pole into a large logarithm containing 
the electron mass. The soft-collinear 
correction to the forward-scattering amplitude is
  \begin{equation}
  \label{TotSoftColl}
     i {\cal A}_{LR}^{{\rm C}\times {\rm SC}}= 
    \frac{16\pi^2\alpha_{ew}^2}{M_W^2}\,
    \left(1-\epsilon\right)\,4\,{\cal I}_{\rm SC},
  \end{equation}
where the integral ${\cal I}_{\rm SC}$, denoting the convolution of 
the zero-distance Green function with soft-collinear emission, 
is the same expression as~\eqref{intSoft}, but cannot be evaluated 
employing the same approximations as for 
the soft corrections. For soft-collinear emission along the electron 
direction, the photon momentum satisfies $n_-\cdot k \sim  \Gamma_W$, 
$n_+ \cdot k\sim \Gamma_W\cdot (m_e/M_W)^2$ and 
$k_\perp \sim m_e \,\Gamma_W/M_W$ with opposite-pointing 
light-like vectors $n_\mp$.
In this case, we parametrise the external momenta according to
$ p_1\equiv \alpha\, n_++\beta\, n_-$ and $p_2\equiv \alpha\, n_-+\beta\, n_+$,
and fix $\alpha$ and $\beta$ in terms of the kinematical variables $s$
and $m_e^2$ through the relations $s=4\,(\alpha+\beta)^2$ and $m_e^2= 
4\,\alpha\,\beta$. The result for the soft-collinear configuration reads
  \begin{equation}
  \label{resSoftColl}
    {\cal I}_{SC}= \,\frac{i}{2}\,\left(\frac{\alpha}{\pi}\right)\, \Gamma(\epsilon)\,
    \left(\frac{M_W}{m_e}\right)^{2\epsilon} (e^{\gamma_E} \mu^2)^\epsilon
    \int_0^\infty\, \!\!\! dk\, \frac{G_{\rm C}^{(0)}(0,0;{\cal E}_W-k)}{k^{1+2\epsilon}}.
  \end{equation}

Summing up the soft and soft-collinear corrections~\eqref{TotSoft} 
and~\eqref{TotSoftColl}
and using the explicit results for the integrals ${\cal I}_{\rm S}$ and 
${\cal I}_{\rm SC}$ \eqref{resSoft} and \eqref{resSoftColl}, we obtain
  \begin{eqnarray}
  \label{TotSoft2}
    {\cal A}_{LR}^{{\rm C} \times [\rm S+SC]}&\equiv& 
    {\cal A}_{LR}^{{\rm C} \times {\rm S}}+{\cal A}_{LR}^{{\rm C} 
    \times {\rm SC}}=
    \frac{32\pi\, \alpha^2_{ew}\,\alpha}{M_W^2}
    \left(1-\epsilon\right) \left[-\frac{\sqrt{\pi}}
    {\epsilon\,\Gamma(1\slash 2 -\epsilon)}
    +\, \Gamma(\epsilon)\,
    \left(\frac{M_W}{m_e}\right)^{2\epsilon}
    \right]\times \nonumber \\
    &&\times\, (e^{\gamma_E} \mu^2)^\epsilon
    \int_0^\infty\, \!\!\! dk\, \frac{G_{\rm C}^{(0)}(0,0;{\cal
        E}_W-k)}{k^{1+2\epsilon}}.
\end{eqnarray}
As expected, the $\epsilon$ pole cancels in the prefactor of~\eqref{TotSoft2};
the infrared sensitivity of the result is reflected in the large
logarithms $\ln\left( M_W\slash m_e\right)$.
\subsection{Production-vertex and hard-collinear corrections}
\label{sec:hard}
We turn now to the radiative corrections to the total cross section
of~\eqref{cor:proc} obtained replacing the LO production 
operator~\eqref{cor:ver} with the NLO expression
  \begin{equation}
    {\cal O}_p^{(1)}= \frac{\pi\, \alpha_{ew}}{M_W^2}\left[
    C_{p,LR}^{(1)}\left(\overline{e}_L \gamma^{[i}n^{j]}e_L\right)+
    C_{p,RL}^{(1)}\left(\overline{e}_R \gamma^{[i}n^{j]}e_R\right)\right]
    \left( \Omega_-^{\dag i} \Omega_+^{\dag j}\right).
  \end{equation}
Here the short-distance one-loop coefficients $C_{p,LR}^{(1)}$ and
$C_{p,RL}^{(1)}$ follow from the matching procedure introduced 
in~\cite{Beneke:2004km,Beneke:2004xd}: 
first, the one-loop $e^-_{L\slash R} e^+_{R\slash L}\to W^- W^+$ 
scattering amplitude is computed at LO in the non-relativistic 
approximation; next, the result is matched with the amplitude obtained 
with the tree-level operator in the effective theory. 
For our purposes, the one-loop coefficient $C_{p,RL}^{(1)}$ is irrelevant, 
since the $e_R^- e_L^+$ helicity configuration 
does not give any LO contribution, and no interference between the
$RL$ and $LR$ configurations arises. As explained in~\cite{Beneke:2007zg}
we need to take into account only the real part of $C_{p,LR}^{(1)}$,
  \begin{equation}
  \label{coeff1loop}
    \mbox{Re}\, C_{p,LR}^{(1)} = \frac{\alpha}{2\pi} \,\mbox{Re} \left[
    \left(-\frac{1}{\epsilon^2}-\frac{3}{2\,\epsilon}\right)
    \left(-\frac{4\,M_W^2}{\mu^2}\right)^{\!-\epsilon} \!\!\!+ 
    c_{p,LR}^{(1,{\rm fin})} \right].
  \end{equation}
The finite part $c_{p,LR}^{(1,{\rm fin})}$ is given explicitely in
Appendix~B.1 of~\cite{Beneke:2007zg}; numerically, we have
$
    \mbox{Re}\, c_{p,LR}^{(1,{\rm fin})} = -10.076
$
for $M_W=80.377$ GeV, $M_Z=91.188$ GeV, top-quark mass $m_t=174.2$ GeV and 
Higgs mass $M_H=115$ GeV. 
The hard production-vertex contributions to the Coulomb-corrected 
forward-scattering amplitude can be readily obtained appending 
twice the one-loop coefficient \eqref{coeff1loop}
to the expression in~\eqref{ampC},
  \begin{equation}
  \label{prodver}
    \mbox{Im} \, {\cal A}_{LR}^{{\rm C}\times {\rm H}} = 
    \frac{16\pi^2 \alpha^2_{ew}}{M_W^2} \, (1-\epsilon)\,
    2\, \mbox{Re} \, C_{p,LR}^{(1)} \,\, \mbox{Im}\, G_{\rm C}^{(0)}(0,0;{\cal E}_W).
  \end{equation}

  \begin{figure}[h]
  \begin{center}
  \includegraphics[width=0.9\textwidth]{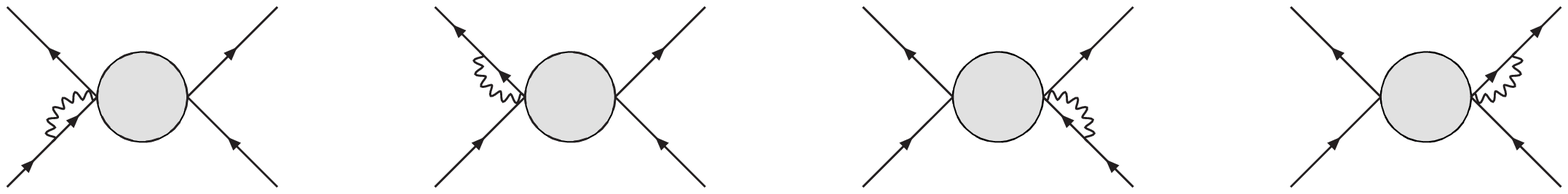}
  \caption{Hard-collinear photon corrections to the all-order Coulomb-corrected 
           forward-scattering amplitude.}
  \label{fig:Collinear}
  \end{center}
  \end{figure}
  Finally, we consider the hard-collinear photon corrections
  illustrated in figure~\ref{fig:Collinear}, associated with momentum
  scalings $k^0\sim M_W$ and $k^2\sim m_e^2$. The collinear-photon
  couplings arise from the SCET Lagrangian, and their couplings to the
  $W$ bosons are encoded in collinear Wilson lines $W_{c_{1/2}}$
  incorporated in the production operators by replacing the electron field
  $e_{c_1,L}$ by $W_{c_1}^{\dagger}e_{c_1,L}$ and analogously 
  for the positron field  $\bar e_{c_2,L}$.  
  Hard-collinear photon corrections to
  the forward-scattering amplitude clearly have a factorised form;
  their contribution can be obtained multiplying \eqref{ampC} by the
  hard-collinear factor derived in~\cite{Beneke:2007zg},
  \begin{eqnarray}
  \label{hcoll}
    {\cal A}_{LR}^{{\rm C}\times {\rm HC}}&=& 
    \frac{16\pi^2 \alpha^2_{ew}}{M_W^2} \, (1-\epsilon)\,G_{\rm C}^{(0)}(0,0;{\cal E}_W) \times\nonumber\\
    &\times& \frac{\alpha}{\pi}\bigg\{
    \frac{1}{\epsilon^2}+\frac{1}{\epsilon}\bigg[-2\ln\left(\frac{m_e}{\mu}\right)
    +\frac{3}{2}\bigg]+2\,\ln^2\left(\frac{m_e}{\mu} \right)
    - 3\ln\left( \frac{m_e}{\mu}\right) + \frac{\pi^2}{12}+3
    \bigg\} .
  \end{eqnarray}

Summing up the production-vertex  and hard-collinear photon corrections \eqref{prodver} and
\eqref{hcoll} we obtain
  \begin{eqnarray}
  \label{hardTOT}
    \mbox{Im} \, {\cal A}_{LR}^{{\rm C}\times [{\rm H}+{\rm HC}]}
    &\equiv& \mbox{Im} \, \left({\cal A}_{LR}^{{\rm C}\times {\rm H}}+  {\cal A}_{LR}^{{\rm C}\times {\rm HC}}\right)
    = 
    \frac{16\pi \,\alpha_{ew}^2\alpha}{M_W^2} \, (1-\epsilon)\,\mbox{Im}\, G_{\rm C}^{(0)}(0,0;{\cal E}_W)
    \nonumber \times\\
    && \hspace*{-2cm}\times\,\bigg\{
    2  \ln\left(\frac{2M_W}{m_e}\! \right)\! 
    \left[\frac{1}{\epsilon}+\ln\left(\frac{2M_W}{m_e} \! \right)
    -\ln\left(\frac{4M_W^2}{\mu^2}\!\right)
      + \frac{3}{2}\right]\!
    +3+\frac{7\pi^2}{12} +\mbox{Re}\, c_{p,LR}^{(1,{\rm fin})} \bigg\}.
  \end{eqnarray}
\subsection{The total cross section}
\label{sec:total}
The combination of ~\eqref{TotSoft2} and~\eqref{hardTOT} represents
our basic expression for 
the higher order contributions to the all-order Coulomb-corrected 
forward-scattering 
$e^-_L e^+_R$ amplitude in the effective theory. In order to better 
exploit the 
structure of the result and show how the cancellation of poles 
in the $\epsilon$
plane takes place, we introduce a modified `$+$' distribution,
  \begin{equation}
    \int_0^\infty\,\!\!\! dk\,\frac{f(k)}{[k]_{a+}}\,\equiv\,
    \int_0^a\,\!\!\! dk\,\frac{f(k)-f(0)}{k}\,+\,\int_a^\infty\,\!\!\!
    dk\,\frac{f(k)}{k},
  \end{equation}
where $a$ is an arbitrary positive real parameter. This parametrisation
allows us to cast the integral containing the zero-distance Coulomb
Green function on the right-hand side of~\eqref{TotSoft2} into
  \begin{equation}
    \mu^{2\epsilon}\,\int_0^\infty\,\!\!\!dk\,\frac{f(k)}
    {k^{1+2\epsilon}}\,=\,
    \left[-\,\frac{1}{2\epsilon}\,+\,\ln\left(\frac{a}{\mu}\right)\,\right]\,
    f(0)
    +\,\int_0^\infty\,\!\!\!dk\,
    \frac{f(k)}{[k]_{a+}}+{\cal O}(\epsilon),
  \end{equation}
with $f(k)\equiv G_{\rm C}^{(0)}(0,0;{\cal E}_W-k)$;
therefore we can prove explicitely that the sum of \eqref{TotSoft2} and \eqref{hardTOT}
is free from $\epsilon$ poles. The full correction 
to the total cross section \eqref{cor:proc} reads
  \begin{eqnarray}
  \label{partonic}
    \overline{\sigma}_{LR}^{\rm C\times[S+H]}\,&\equiv&\,
    \frac{1}{27\,s}\mbox{Im}\,
    \left( {\cal A}_{LR}^{{\rm C}\times[{\rm S}+{\rm SC}]}\,
    +\,{\cal A}_{LR}^{{\rm C}\times[{\rm H}+{\rm HC}]}\right)\nonumber\\
    &=& \,\frac{16\,\pi\,\alpha_{ew}^2\,\alpha}{27\,s\,M_W^2}\,
    \Bigg\{4\ln\left(\frac{2 M_W}{m_e}\right)\mbox{Im}\,
    \int_0^\infty\,\!\!\! dk\,
   \frac{G_{\rm C}^{(0)}(0,0;{\cal E}_W-k)}{[k]_{a+}}\\
    &&+\,
   \Bigg[\Bigg( 3\! - 4\,\ln\left(\frac{M_W}{a}\right)\Bigg)\,
    \ln\left(\frac{2 M_W}{m_e}\right) + 3+\frac{\pi^2}{4}
    +\mbox{Re}\, c_{p,LR}^{(1,{\rm fin})} \Bigg]
    \mbox{Im} \, G_{\rm C}^{(0)}(0,0;{\cal E}_W)
    \Bigg\}.\nonumber
  \end{eqnarray}
Now, after all poles in $\epsilon$ have cancelled, 
$G_{\rm C}^{(0)}(0,0;{\cal E}_W)$ can be taken to be the
four-dimensional expression given in ~\eqref{coulombGF}.
The result of~\cite{Beneke:2007zg} for the NLO correction to the 
total cross section at the partonic level can be readily re-derived 
inserting the zero-Coulomb exchange term 
of~\eqref{coulombGF} into~\eqref{partonic}, and observing that
  \begin{equation}
    \mbox{Im} \, \int_0^\infty\,\!\!\! dk\,\frac{1}{[k]_{a+}}\,
    \sqrt{- \frac{{\cal E}_W-k}{M_W}}
    =\mbox{Im} \bigg\{\sqrt{- \frac{{\cal E}_W}{M_W}}\left[ 
    \ln\left( -\frac{4 {\cal E}_W}{a}\right)-2\right]\bigg\}.
  \end{equation}

The new N$^{3\slash2}$LO${}^{\rm EFT}$ correction follows from
inserting the one-Coulomb exchange term of~\eqref{coulombGF} 
into~\eqref{partonic} for the total cross section; 
after performing explicitely
the $k$ integration,
  \begin{equation}
  \label{integrate}
    \mbox{Im}  \int_0^\infty\,\!\!\!\! dk \frac{1}{[k]_{a+}}
    \ln\left( \!-\frac{4M_W({\cal E}_W-k)}{\mu^2} \!\right)
    =\mbox{Im} \left[ \!\frac{1}{2}\ln^2\left(-\frac{{\cal E}_W}{M_W}\right) 
    - \ln\left(-\frac{{\cal E}_W}{M_W}
    \right)\ln\left(\frac{a}{M_W}\right)\! \right],
  \end{equation}
we get
  \begin{eqnarray}
    \Delta \overline{\sigma}_{LR}^{\rm C\times[S+H]}&=& 
    \frac{4\,\alpha_{ew}^2\alpha^2}{27\,s}
    \mbox{Im}\bigg\{\!\! -\frac{1}{2}\ln\bigg( \!\! -\frac{{\cal E}_W}{M_W}\bigg)\bigg[
    2\ln\bigg( \!\! -\frac{{\cal
        E}_W}{M_W}\bigg)\ln\bigg(\frac{2M_W}{m_e}\! 
    \bigg)\nonumber\\
    &&+\,3\ln\bigg(\frac{2M_W}{m_e}\bigg)+3+\frac{\pi^2}{4}+
    \mbox{Re}\, c_{p,LR}^{(1), {\rm fin}} \bigg]\bigg\}.
  \end{eqnarray}
The dependence on the $a$ regulator has obviously cancelled; the result is
finite in the $\epsilon$ plane, but the infrared sensitivity of the cross
section is reflected in the large logarithms $\ln(2M_W\slash m_e)$.
In the next section we show how to absorb the large logarithms in
the electron distribution function.

\subsection{Initial-state radiation}
\label{sec:isr}
An accurate prediction for the total cross section of~\eqref{cor:proc}
requires to resum the collinear logarithms from initial-state radiation (ISR).
Here we apply the strategy outlined in~\cite{Beneke:2007zg}, employing
the electron structure function $\Gamma_{ee}^{\rm LL}$ provided
in~\cite{Skrzypek:1992vk,Beenakker:1996kt}. The ISR 
resummed cross section for a given helicity configuration $h$ is
  \begin{equation}
  \label{convolute}
    \sigma_h(s)\equiv\int_0^1 dx_1\int_0^1 dx_2\, \Gamma_{ee}^{\rm LL}(x_1)\,
    \Gamma_{ee}^{\rm LL}(x_2)\, \hat{\sigma}_{h}(x_1x_2s),
  \end{equation}
where the partonic cross section $\hat{\sigma}_{h}$ 
denotes the sum of the Born result and the 
higher-order correction,  
$\hat{\sigma}_{h}=\sigma_{h}^{\rm Born} +\hat{\sigma}_{h}^{\rm h.o.}$,
with logarithms of the electron mass subtracted such that 
expansion of (\ref{convolute}) reproduces the NLO cross section.

Therefore, we construct $\Delta \hat{\sigma}_{LR}^{\rm{C}\times [\rm{S+H}]}$
starting from $\Delta \overline{\sigma}_{LR}^{\rm{C}\times [\rm{S+H}]}$ 
of~\eqref{partonic} and performing a subtraction,
  \begin{equation}
  \label{sub}
    \Delta \hat{\sigma}_{LR}^{\rm{C}\times [\rm{S+H}]}(s)\equiv  
    \Delta \overline{\sigma}_{LR}^{\rm {C}\times [\rm{S+H}]}(s)
    -2\int_0^1 dx \,\Gamma_{ee}^{{\rm LL},(1)}(x)\sigma_{LR}^{\rm C}(xs),
  \end{equation}
where $\Gamma_{ee}^{{\rm LL},(1)}$ is the ${\cal O}(\alpha)$ term in the
expansion of the conventional structure function provided 
in~\cite{Skrzypek:1992vk,Beenakker:1996kt}, evaluated at $\sqrt{s}=2M_W$,
and $\sigma_{LR}^{\rm C}$ is related to the all-order Coulomb-corrected 
forward-scattering amplitude~\eqref{ampC} through the usual relation 
$\sigma_{LR}^{\rm C}= \mbox{Im} \,{\cal A}_{LR}^{\rm C}\slash (27\,s)$.
Note that the zero-distance Coulomb Green function appearing in~\eqref{ampC}
has now to be evaluated with the replacement $E\to E-M_W(1-x)$.
Using $\Gamma_{ee}^{{\rm LL},(1)}$ in the limit $x\to 1$,
  \begin{equation}
    \Gamma_{ee}^{{\rm LL},(1)}(x)\to \frac{\alpha}{2\pi}\left[2\,\ln\left(
    \frac{2 M_W}{m_e}\right)-1\right]\left\{ \frac{2}{[1-x]_+} +\frac{3}{2}
    \delta(1-x)\right\},
  \end{equation}
we can write the subtraction term appearing in~\eqref{sub} as
  \begin{eqnarray}
  \label{subtractionTerm}
    -2\int_0^1 dx \,\Gamma_{ee}^{{\rm LL},(1)}(x)\sigma_{LR}^{\rm C}(xs)
    &=&-\frac{16\,\pi\alpha_{ew}^2\alpha}{27\,s\,M_W^2} \left[2\,\ln\left(
    \frac{2 M_W}{m_e}\right)-1\right]\times\nonumber\\
    && \hspace*{-2cm} 
  \times \,\mbox{Im}\bigg\{ \frac{3}{2} G_{\rm C}^{(0)}(0,0;{\cal E}_W)
    +2\,\int_0^\infty\, \!\!\! dk\, 
    \frac{G_{\rm C}^{(0)}(0,0;{\cal E}_W-k)}{[k]_{M_{W}+}}\bigg\}.
  \end{eqnarray}
To obtain this result, a term involving the integral 
$\mbox{Im}[\,\int_{M_W}^\infty dk \,G_{\rm C}^{(0)}(0,0;{\cal
    E}_W-k)/k\,]$, which is suppressed by a power of 
${\cal E}_W/M_W$ and corresponds to hard-collinear initial-state radiation, 
has been dropped.
Replacing~\eqref{subtractionTerm} and~\eqref{partonic} in~\eqref{sub}
and choosing $a=M_W$, we finally obtain
  \begin{equation}
    \hat{\sigma}_{LR}^C(s)= \frac{16\pi\alpha_{ew}^2\alpha}{27sM_W^2}\bigg[
    \bigg( \frac{9}{2}+\frac{\pi^2}{4}+\mbox{Re}\, c_{p,LR}^{(1),{\rm fin}}
    \bigg)\,\mbox{Im}\, G_{\rm C}^{(0)}(0,0;{\cal E}_W)
    + 2 \,\mbox{Im}\, \int_0^\infty \!\!\! dk \,
    \frac{G_{\rm C}^{(0)}(0,0;{\cal E}_W-k)}{[k]_{M_W+}}
    \bigg],
  \end{equation}
where the dependence on the large logarithms $\ln(2M_W\slash m_e)$ has
cancelled out.

Using~\eqref{integrate} for the one-Coulomb exchange term in~\eqref{coulombGF},
we get the N$^{3\slash 2}$LO$^{\text {EFT}}$ contribution to the 
total cross section to be convoluted
with the electron structure functions in~\eqref{convolute},
  \begin{equation}
\label{eq:sigma-three-half}
\Delta\hat{\sigma}_{LR}^{\rm{C}\times [\rm{S+H}]} =
    -\frac{\alpha_{ew}^2\,\alpha^2}{27\,s}
    \bigg\{ \bigg( 9+\frac{\pi^2}{2}+2\,\mbox{Re}\, c_{p,LR}^{(1),{\rm fin}}
    \bigg)\mbox{Im}\bigg[\ln\left(-\frac{{\cal E}_W}{M_W} \right)
    \bigg]
    +2\,\mbox{Im}\bigg[\ln^2\left(-\frac{{\cal E}_W}{M_W} \right)
    \bigg] \bigg\}.
  \end{equation}
The result has been cross-checked through a direct integration
of the one-Coulomb exchange terms. 


\subsection{Corrections to the Coulomb potential}
\label{sec:bubble}

N$^{3\slash 2}$LO$^{\text {EFT}}$ contributions also arise from 
semi-soft and hard corrections to 
the Coulomb force between the two $W$s. 
The hard corrections are closely related to the renormalisation of the 
electromagnetic coupling in the Coulomb potential; 
they are discussed in detail in 
appendix~\ref{sec:hard-coulomb}. 

\begin{figure}[ht]
  \begin{center}
  \includegraphics[width=\textwidth]{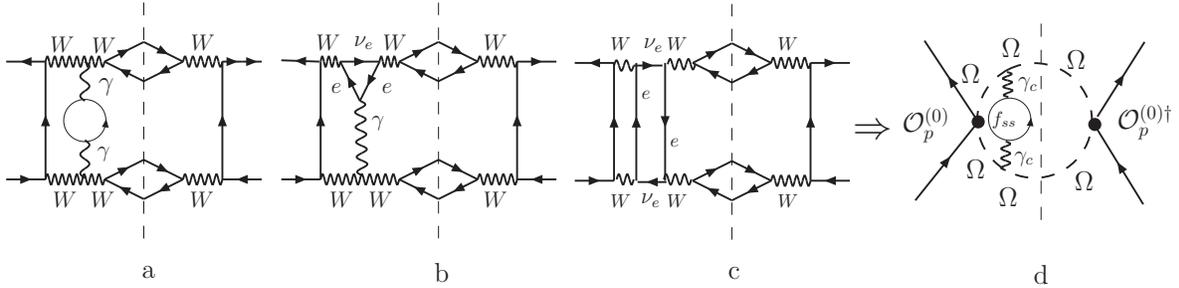}
  \caption{a-c: Sample diagrams in the full SM that potentially give rise to corrections from semi-soft fermions. 
  Only the insertion of a  semi-soft fermion bubble (a) contributes at N$^{3\slash 2}$LO and leads to the EFT diagram shown in d.}
  \label{fig:semi-soft}
  \end{center}
  \end{figure}

As a first example for semi-soft modes,  consider corrections from light
fermions that consist of the insertion of a fermion bubble into a
photon propagator exchanged between two $W$-lines as shown in
figure~\ref{fig:semi-soft}a but also vertex corrections
(figure~\ref{fig:semi-soft}b) and box corrections
(figure~\ref{fig:semi-soft}c). For a potential photon the only
correction at order N$^{3\slash 2}$LO comes from the bubble insertion where
the fermion is semi-soft, leading to the effective theory diagram
shown in figure~\ref{fig:semi-soft}d.  Since the semi-soft fermion
propagator $1/\!{\not \!k_{ss}}$ counts as $\delta^{-1/2}$ and the
loop measure as $d^4k_{ss} \sim \delta^2$ the fermion loop insertion
scales as $\alpha \delta^{1}\sim \delta^2$. Taking the additional
potential photon propagator $\sim \delta^{-1}$ into account this
diagram is suppressed by a factor $\delta$ compared to the pure 
single-Coulomb exchange and is therefore a N$^{3\slash 2}$LO correction to
the forward-scattering amplitude.  The triangle diagrams and box
diagrams with semi-soft fermions do not contribute at order
N$^{3\slash 2}$LO as can be seen from a power-counting argument. For
instance, in the triangle diagram of figure~\ref{fig:semi-soft}b the
semi-soft loop momentum can be neglected compared to the $W$-mass so
the neutrino propagator contains no momentum dependence. By the same
arguments as before the diagram is of order $\delta^2$ compared to the
single-Coulomb exchange and does not have to be considered.

\begin{figure}[ht]
  \begin{center}
  \includegraphics[width=\textwidth]{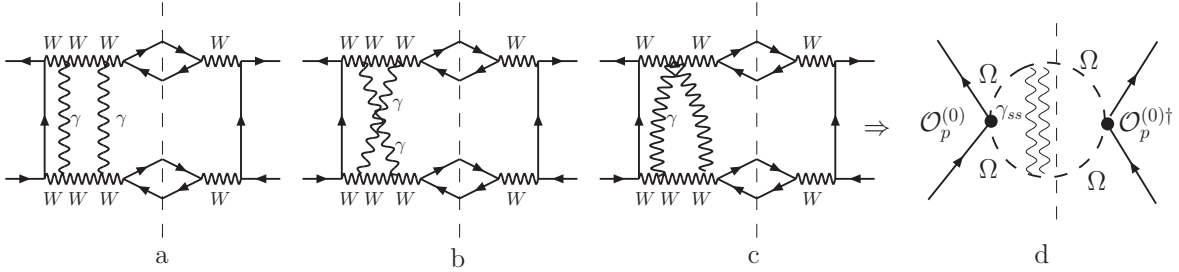}
  \caption{a-c: Sample diagrams in the full SM that potentially give rise to corrections from semi-soft photons. A sample diagram in the EFT is shown in d. 
    As discussed in the text these diagrams do not contribute at N$^{3\slash 2}$LO.}
  \label{fig:semi-soft-box}
  \end{center}
  \end{figure}

Another N$^{3\slash 2}$LO correction potentially arises from  box diagrams 
with exchange of two semi-soft photons.  Example diagrams in the full theory
are shown in figure~\ref{fig:semi-soft-box}a-c, and 
a representative diagram in the effective theory is shown in 
figure~\ref{fig:semi-soft-box}d.
After interacting with a semi-soft photon with momentum $k_{ss}$, 
the $W$ propagator turns into $
1/(2M_W k_{ss,0})\sim \delta^{-1/2}$.
 Note that $k_{ss,0}\sim M_W\sqrt \delta \gg \Gamma_W$  so that the decay width can be treated perturbatively in the semi-soft region.
Since the  photon propagators are given by $1/k_{ss}^2\sim \delta^{-1}$, the semi-soft subgraph
counts as $\alpha^2 d^{4}k_{ss} (1/k_{ss}^2)^2 1/(2k_{ss,0})^2\sim \alpha^2 \delta^{-1}$. 
The potential sub-loop of the box diagram figure~\ref{fig:semi-soft-box}d
  has two $\Omega$ propagators so 
it scales as $\delta^{5/2}\delta^{-2}$. 
Therefore the contribution of 
these diagrams would again be suppressed by $\delta^{3\slash 2}$ compared to the leading order. However, at leading order  in the non-relativistic expansion
the diagrams shown in figure~\ref{fig:semi-soft-box}a  and 
\ref{fig:semi-soft-box}b cancel each other.
For this it is essential that the width in the propagator can be treated perturbatively in the semi-soft region and is not resummed in the propagator.
Diagrams that contain a quartic  vertex like~\ref{fig:semi-soft-box}c  
contribute only at higher orders because they  miss at least one
factor  $1/k_{ss,0}$ from a $W$-propagator compared to figure~\ref{fig:semi-soft-box}a/b. Therefore there are no contributions from semi-soft photons at the order we consider here.
 In  QED this is  well known from explicit
calculations of the Coulomb potential~\cite{Gupta:1981pd,Titard:1993nn}.

  The only corrections to be calculated are therefore the semi-soft
  fermion bubble insertion in the single-Coulomb exchange diagram
  shown in figure~\ref{fig:semi-soft}d and the hard corrections
  discussed in appendix~\ref{sec:hard-coulomb}.  For convenience we
  will perform the calculation in the $\alpha(M_Z)$ scheme (see e.g.
  \cite{Dittmaier:2001ay,Denner:1991kt}) 
  and convert the result afterwards to the
  $G_\mu$ scheme used in~\cite{Beneke:2007zg}.  The hard corrections
  to the forward-scattering amplitude are proportional to the single
  Coulomb exchange, hence
\begin{equation}
  \Delta
  \mathcal{A}^{\text{NLO-C}}_{LR}|_{\alpha(M_Z)}=
  \Delta \mathcal{A}^{\text{ss}}_{LR} + 
  \mathcal{A}^{\text{C1}}_{LR} \; 
\delta_{\text{hard}}^{\alpha(M_Z)},
\end{equation} 
where $ \Delta \mathcal{A}^{\text{ss}}_{LR}$ is the 
unrenormalised semi-soft amplitude shown in figure~\ref{fig:semi-soft}d, and 
$\mathcal{A}^{\text{C1}}_{LR}$ the single-Coulomb correction to 
the forward-scattering amplitude  included in~\eqref{ampC}.
The hard correction is given in~\eqref{eq:delta-mz} 
and involves the contribution of the light fermions (leptons and
quarks except top) to the
vacuum polarisation evaluated at $M_Z$:
\begin{equation}
\label{eq:bubble-counter}
\delta_{\text{hard}}^{\alpha(M_Z)}=  
-\frac{\re \,\Pi_{f\neq t}^{AA}(M_Z^2)}{M_Z^2}
= \sum_f C_f Q_f^2 \,\,2\left( \frac{\alpha}{\pi} \right)
     e^{\gamma_E \epsilon} \,\Gamma(\epsilon)  \,\mbox{Re} \left[ 
     \left( -\frac{M_Z^2}{\mu^2}\right)^{-\epsilon}\,\right] 
     \frac{\Gamma^2(2-\epsilon)}
     {\Gamma(4-2\epsilon)}.
 \end{equation}
Here $C_f=1$ for the leptons and $C_f=N_c=3$ for the quarks, and $Q_f$
is the electric charge of $f$ in units of $e$, such that 
$\sum_f C_f Q_f^2=20/3$. The total correction to the
  forward-scattering amplitude from the semi-soft bubble and the 
hard correction, at all orders in $\epsilon$, is given by
\begin{eqnarray}
  \Delta \mathcal{A}^{\text{NLO-C}}_{LR}|_{\alpha(M_Z)}
   &=&
  -\sum_f C_f Q_f^2 \,\frac{\alpha_{ew}^2\alpha^2}{2\pi}\,
    e^{3\gamma_E \epsilon}\,(1-\epsilon)\,\frac{1}{\epsilon} \frac{
    \Gamma(\epsilon)
    \Gamma^2(2-\epsilon)}{\Gamma(4-2\epsilon)\Gamma(3\slash
    2-\epsilon)} \,\times 
\nonumber \\
    && \hspace*{-2.2cm}\times\,\bigg[
    \left(\frac{M_W}{\mu} \right)^{\!-3\epsilon} 
    \left(- \frac{{\cal E}_W}{\mu}\right)^{\!-3\epsilon}
    \frac{\Gamma(3\epsilon)\Gamma(1\slash 2-2\epsilon)\Gamma^2(1\slash
      2+2\epsilon)}{\Gamma(4\epsilon)}
\nonumber\\
    && \hspace*{-1.5cm} - 2  \,\left(\frac{M_W}{\mu} \right)^{\!-2\epsilon} 
    \left( - \frac{{\cal E}_W}{\mu}\right)^{\!-2\epsilon}
    \mbox{Re} \left[\left(-\frac{M_Z^2}{\mu^2}\right)^{\!-\epsilon}\,\right]
        \Gamma(1\slash 2-\epsilon) \Gamma^2(1\slash 2+\epsilon)
    \bigg].
  \label{bubble}
  \end{eqnarray}
Expanding \eqref{bubble} in $\epsilon$ one gets
  \begin{equation}
    \Delta\sigma_{LR}^{\rm NLO-C}|_{\alpha(M_Z)}= -
    \frac{\alpha_{ew}^2\alpha^2}{81 s} 
     \sum_f C_f Q_f^2\left\{
     4\ln\!\left(\frac{2M_W}{M_Z}\right)
     \mbox{Im}\left[\ln\!\left( -\frac{{\cal E}_W}{M_W}\right)\right]
     + \mbox{Im}\left[\ln^2\!\left( -\frac{{\cal E}_W}{M_W}\right)\right]
    \right\}.
  \end{equation} 
While this result is valid in the $\alpha(M_Z)$ scheme,  the 
numerical results in~\cite{Beneke:2007zg} were given in the
 $G_\mu$ scheme.
According to the discussion in appendix~\ref{sec:hard-coulomb}
the result~\eqref{bubble} can be converted to the latter scheme 
by adding another term:
  \begin{equation}
\label{eq:bubble-gmu}
\begin{aligned}
    \Delta\sigma_{LR}^{\text{NLO-C}}
    &= \Delta\sigma_{LR}^{\text{NLO-C}}|_{\alpha(M_Z)}+
    \delta_{\alpha(M_Z)\to G_\mu}
\Delta\sigma^{\text{C1}}_{LR}\\
&=  \Delta\sigma_{LR}^{\text{NLO-C}}|_{\alpha(M_Z)}
- \frac{2\pi \alpha_{ew}^2\alpha }{27 s} \,
\delta_{\alpha(M_Z)\to G_\mu}\,
\mbox{Im}\left[ \ln\!\left( -\frac{{\cal E}_W}{M_W}\right)\right],
\end{aligned}
\end{equation}
where the single-Coulomb exchange cross-section $ \Delta
\sigma^{\text{C1}}_{LR}$ is the order $\alpha$ correction
in~\eqref{CoulombCS}. 
The result for the finite conversion factor $\delta_{\alpha(M_Z)\to G_\mu}$
is given in~\eqref{eq:delta-mz-gmu}
in appendix~\ref{sec:hard-coulomb}.
For the same input parameters as used for the hard-matching coefficient below~\eqref{coeff1loop} the numerical value  is
given by $\delta_{\alpha(M_Z)\to G_\mu}=4.103 \alpha$. 

\subsection{Decay corrections}
\label{sec:decay}
In~\eqref{eq:sigma-three-half} the radiative corrections to
the imaginary part of the forward-scattering amplitude are multiplied
by the leading-order branching fraction product $
\text{Br}^{(0)}(W^-\to \mu^- \bar{\nu}_\mu) \text{Br}^{(0)}(W^+ \to u
\bar{d}\,)=1/27$ to extract the correction to the flavour-specific
four-fermion production cross section.  This treatment can be shown to
correspond to cutting the string of fermion bubbles implicitly
contained in the resummed propagator and selecting only those cuts
that contribute to the desired four-fermion final
state~\cite{Beneke:2007zg}.  In addition to the effects described correctly
 by~\eqref{eq:sigma-three-half} there are flavour-specific
radiative corrections to the decay stage contributing at N$^{3\slash
  2}$LO$^{\text {EFT}}$. These arise from diagrams in
the full theory such as shown in figure~\ref{fig:NLOcount}e, 
and the cuts corresponding to real emission from the final state.

These flavour-specific corrections
 are included in the EFT calculation by adding the term
\begin{equation}
  \Delta\sigma_{LR}^{\rm C\times decay}=
\left(\frac{\Gamma^{(1,ew)}_{\mu^-\bar\nu_\mu}}{{
\Gamma^{(0)}_{\mu^-\bar\nu_\mu} }}+
\frac{\Gamma_{u\bar d}^{(1,ew)}}{\Gamma_{u\bar d}^{(0)} }
\right) \Delta \sigma^{\text{C1}}_{LR},
\label{eq:delta-decay}
\end{equation}
where the one-loop electroweak corrections to the
partial decay-widths, $\Gamma^{(1,ew)}_{\mu^-\bar\nu_\mu}$ and
$\Gamma_{u\bar d}^{(1,ew)}$, can be found in~\cite{Beneke:2007zg}. Note
that the numerical predictions presented in~\cite{Beneke:2007zg}
already include the 2-loop QCD corrections to the hadronic decay-width
multiplied by the full NLO electroweak cross section (including
the single-Coulomb exchange). Therefore QCD corrections do not have to
be considered in this work.

\subsection{Residue corrections to the $W$-propagators}
\label{sec:residue}

In the proper effective-theory treatment of the computation of
matching coefficients one has to include a factor $\sqrt{2M_W}(\varpi
R_{hW})^{-1/2}$ for each external $\Omega$
line~\cite{Beneke:2007zg,Beneke:2004km}.  Here $\varpi$ accounts for
the normalisation of non-relativistic fields and $R_{hW}$ is the hard
contribution to the LSZ residue of the $W$ propagator.  Introducing an
expansion of the transverse self-energy around $M_W^2$ and in the
number of loops,
\begin{equation}
\Pi_T^W(k^2) = M_W^2 \sum_{m,n}  \delta^n \, \Pi^{(m,n)},
\label{eq:Pihard}
\end{equation}
with $\delta=(k^2-M_W^2)/M_W^2$ and $m$ denoting the loop order one 
has to one-loop accuracy~\cite{Beneke:2004km}
\begin{equation}
  R^{-1}_{hW}=1-\Pi^{(1,1)}.
\end{equation}
Using the on-shell scheme for field renormalisation where 
$\re\,\Pi^{(1,1)}=0$, the residue correction is purely imaginary, 
$R_{hW}^{-1}=1-i \,\im \,\Pi^{(1,1)}=1+i \Gamma_W^{(0)}/M_W$ 
with $\Gamma_W^{(0)}$ the tree-level on-shell width.
In the EFT treatment these factors reproduce the 
expansion of the full renormalised (transverse) $W$-propagator
\begin{equation}
\label{eq:expand-propagator}
  P(k) = \frac{i}{k^2-M_W^2-\Pi_T^W(k^2) }=  
\frac{i}{2 M_W \left(r_0-\frac{\vec{r}^{\,2}}{2M_W}
+ \frac{i \Gamma_W^{(0)}}{2}\right)}
\left(1+\Pi^{(1,1)}+\varpi^{(1)}\right)+ \dots,
\end{equation}
where the NLO contribution to the non-relativistic normalisation
factor, $\varpi^{(1)}$, can be found in~\cite{Beneke:2007zg}, but 
is not needed here since in the full NLO calculation in the
fixed-width or complex-mass scheme no non-relativistic expansion is 
performed.

At order N$^{3\slash 2}$LO the residue corrections have to be included
in the single-Coulomb exchange diagram. This corresponds 
to including the proper
factors of $\sqrt{R_{hW}}$ in the NLO matching coefficient
$C_{p,LR}^{(1)}$ of the production operator in the calculation 
in section~\ref{sec:hard} and the $WW\gamma$ vertex in 
appendix~\ref{sec:hard-coulomb}. 
In order to separate the effects corresponding to 
genuine loop diagrams in the standard loop expansion from those 
included in  a tree-calculation using a fixed-width prescription, 
in~\cite{Beneke:2007zg} and this paper 
we departed from the matching  procedure 
sketched above and did not include the factors $\varpi R_{hW}$ 
in the computation of the hard matching coefficients. 
Therefore they have to be discussed separately,
paying some attention to isolate those contributions that are not
included in a fixed-order NLO 
calculation in the fixed-width or complex-mass scheme.

\begin{figure}[t]
  \begin{center}
  \includegraphics[width=0.60\textwidth]{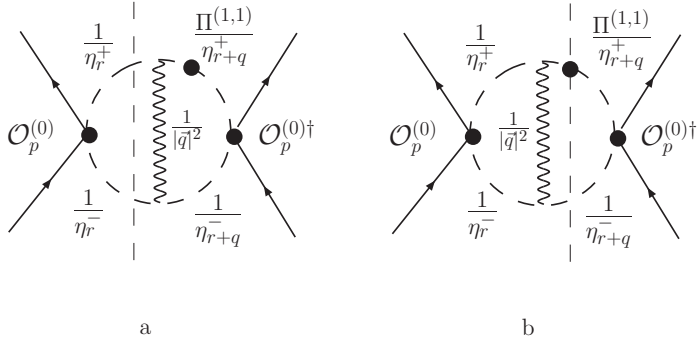}
  \caption{Cut EFT diagrams contributing to the residue corrections to the
  imaginary part of the forward-scattering amplitude. The $\Omega$ 
  propagator with a dot indicates the insertion of a $\Pi^{(1,1)}$ 
  correction factor coming from the expansion of the propagator around
  the pole~\eqref{eq:expand-propagator}.
  The propagators are defined in~\eqref{eq:def-eta}. Analogous
  diagrams with insertions at the other three $\Omega$ propagators are 
  not shown.} 
  \label{fig:residue}
  \end{center}
  \end{figure}

The factor $\varpi$ is solely due to the use of a 
non-relativistic propagator. Since no kinematic expansion is performed
in the calculation in the fixed-width or 
complex-mass scheme, all corrections of this kind are already 
included in~\cite{Denner:2005es,Denner:2005fg} and do not have to be 
considered here. 
This leaves the corrections from the derivative of the self-energy, $\Pi^{(1,1)}$, in~\eqref{eq:expand-propagator}.
As shown in figure~\ref{fig:residue} these corrections contribute in two different ways to the imaginary part of the forward-scattering amplitude. 
In this figure we indicated the propagators of the non-relativistic $W$s by
the abbreviations 
\begin{equation}
\label{eq:def-eta}
 \eta^{-}_r=r^0-\frac{\vec{r}^{\,2}}{2 M_W}+i\frac{\Gamma^{(0)}_W}{2},
\qquad
 \eta^{+}_r=E-r^0-\frac{\vec{r}^{\,2}}{2
 M_W}+i \frac{\Gamma^{(0)}_W}{2}.
\end{equation}

The cut through the propagator in figure~\ref{fig:residue}a, $2\,\im[1/\eta
\,]=-\Gamma_W^{(0)}/|\eta|^2$, is interpreted as a cut through 
a self-energy insertion implicitly contained in the resummed propagator.
To select the flavour-specific final state, the total decay width
in the numerator has to be replaced by the (leading order) partial
decay width $\Gamma^{(0)}_{\mu^-\bar \nu_\mu}$ or 
$\Gamma^{(0)}_{u\bar d}$~\cite{Beneke:2007zg}. 
This diagram therefore corresponds to a single-Coulomb photon exchange diagram 
in the full theory with the insertion of a
$W$ self-energy, expanded to first order around the mass-shell. These
contributions are not included in the fixed-order NLO calculation in
the fixed-width or 
complex-mass scheme where only the self-energy evaluated at
$M_W^2$ is resummed in the propagator~\cite{Denner:2005fg}.  In
contrast the diagram in figure~\ref{fig:residue}b includes a
cut through a propagator modified by the factor of 
$\Pi^{(1,1)}=-i \Gamma_W^{(0)}/M_W$. The explicit expression  is given by 
\begin{equation}
\label{eq:cut-pi}
2\,\im\left[\frac{\Pi^{(1,1)}}{\eta}\right]
=-\frac{\Gamma_W^{(0)}}{M_W}\left(r_0-\frac{\vec r^{\,2}}{2M_W}\right)
\frac{1}{|\eta|^2} .
\end{equation} 
This contribution  corresponds to 
a full-theory diagram with single-Coulomb exchange and kinematical correction 
from the expansion of a  decay matrix element  around
the mass-shell. 
This interpretation is confirmed by inserting the one-loop result
 $\mbox{Im} \,\Pi_T^W(k^2) =
-k^2 \Gamma_W^{(0)}/M_W \,\theta(k^2)$ for the decay into massless fermions
into the resummed propagator on the left-hand side
of~\eqref{eq:expand-propagator}:
\begin{equation}
\im[-i P(k) ]= \frac{- M_W
\left(k^2 \Gamma_W^{(0)}/M_W^2\right)}
{\left(k^2-M_W^2 \right)^2+M_W^2
\left(k^2 \Gamma_W^{(0)}/M_W^2\right)^2}
+O\left(1\right).
\end{equation}
A fixed-width prescription as employed in the complex-mass scheme
corresponds to replacing
$k^2 \Gamma_W^{(0)}/M_W^2$ by $\Gamma_W^{(0)}$ in
the denominator, but not in the numerator, where the
factor of $k^2$ arises from the integration over the
two-particle phase space of the $W$ decay products.
Decomposing the $W$-momentum as $k=M_Wv + r$ with $v=(1,\vec 0\,)$ and 
a potential residual momentum $r$  shows that the $\mathcal{O}(\delta)$
correction from expanding the factor $k^2$ in the numerator
is indeed given by~\eqref{eq:cut-pi} (up to the usual normalisation 
factor $2M_W$).
Since in the fixed-order calculation of four-fermion
production the decays are treated without kinematical approximations,
this term does not correspond to a genuine NNLO contribution and  
must not be included here.

The only correction that has to be included therefore 
are four cut diagrams  of the form of figure~\ref{fig:residue}a
 with the residue correction inserted at the
different $W$-propagators.  In the following we use again the 
full on-shell width $\Gamma_W$ rather than the tree-expression 
$\Gamma_W^{(0)}$.
Shifting the loop momenta in some of the diagrams,
the sum of the four terms  can be brought to the form
\begin{eqnarray}
  \Delta \sigma_{LR}^{{\rm C}\times \text{res}}
&=&- \frac{32 (1-\epsilon) \pi^3 \alpha_{ew}^2 \alpha}{27 M_W^2 s}
\,\tilde \mu^{4\epsilon}\!
\int\!\frac{d^d r}{(2 \pi)^d}\int \!\frac{d^d q}{(2\pi)^d}\, 
2\,\im\left[\frac{1}{\eta^+_{r}}\right]
2\,\im\left[\frac{1}{\eta^-_{r}}\right]
2\,\re\left[\frac{i}{\vec q^{\,2}}\frac{2\Pi^{(1,1)}}{
 \eta^{+}_{r+q}\eta^{-}_{r+q}}\right] \nonumber\\
&=& -\frac{32(1-\epsilon) \pi^2 \alpha_{ew}^2 \Gamma_W}{27 M_W^3 s} 
\times  \nonumber \\
&& \times \,\Biggl[\re \,G_{C1}^{(0)}(0,0,\mathcal{E}_W) 
\left.-4\pi\alpha \tilde \mu^{4\epsilon}
\int\frac{d^d r}{(2 \pi)^d}\int \frac{d^d q}{(2\pi)^d}  \frac{1}{\vec q^{\,2}}
\frac{1}{\eta^+_{r}\eta^-_r (\eta^+_{r+q})^* (\eta_{r+q}^-)^* }
\right] .
\label{rescorr}
\end{eqnarray}
Here $G^{(0)}_{\rm C1}$ denotes the single-Coulomb exchange term in the
Coulomb Green function~\eqref{coulombGF}. 
Terms that vanish upon performing the $q_0$ and $r_0$ integrations by 
closing the integration contour in the upper half-plane have been dropped.
The explicit computation of~\eqref{rescorr} leads to
\begin{equation}
\label{eq:delta-residue}
\Delta \sigma_{LR}^{{\rm C}\times {\rm res}}=\frac{4 \pi \alpha_{ew}^2
  \alpha }{27 s} \frac{\Gamma_W }{M_W} 
  \ln \left[ \frac{2 \,|{\cal E}_W| 
  \left( \re \,{\cal E}_W+|{\cal E}_W|\right)}{\Gamma_W^2}\right] \, .
\end{equation}


\section{Numerical analysis}
\label{sec:nums}

We can now assemble the N$^{3\slash 2}$LO
contribution to the total
cross section of the scattering process $e^-e^+\to
\mu^-\overline{\nu}_\mu u\overline{d}+X$, given by the sum of the 
several corrections computed in section~\ref{sec:eval},
\begin{equation}
\label{eq:three-half-parton}
  \hat \sigma_{LR}^{(3\slash 2)}=
 \Delta\hat \sigma_{LR}^{{\rm C}\times[\rm{S+H}]}
   +\Delta\sigma_{LR}^{\text{NLO-C}}
   +\Delta\sigma_{LR}^{\rm{C}\times \text{decay}}
   +\Delta\sigma_{LR}^{\rm{C}\times \text{res}}  
   +\Delta\sigma_{LR}^{\rm{C3}},
\end{equation}
which is to be inserted into the convolution with the 
electron structure functions
in~\eqref{convolute}.  Recall that this refers to the $e^-_L e^+_R$
helicity state while there are no radiative
corrections to the other helicity combinations from NNLO SM diagrams
that contribute at N$^{3\slash 2}$LO in the EFT power counting. 
As input parameters for the masses of the gauge bosons and the $W$
width we use
  \begin{equation}
    M_W= 80.377\, \text{GeV},\qquad \Gamma_W= 2.09201\, \text{GeV}, 
    \qquad M_Z=91.188\, \text{GeV}.
  \end{equation}
In addition, we employ for the masses of the electron, the top quark 
and the Higgs boson
  \begin{equation}
    m_e= 0.51099892\, \text{MeV},\qquad m_t=174.2\,\text{GeV},\qquad
    M_H=115\,\text{GeV},
  \end{equation}
and we extract the fine-structure constant $\alpha$ from the
 relation $\alpha=\sqrt{2}G_\mu M_W^2 s^2_w\slash \pi$, where the 
Fermi-coupling constant is $G_\mu=1.16637\cdot 10^{-5}
\,\text{GeV}^{-2}$ and the cosine of the weak-mixing angle reads 
$c_w= M_W\slash M_Z$.
  \begin{table}[p]
  \begin{center}
  \begin{tabular}{|c||c||c|c|c|c|c||c|}
    \hline&
    \multicolumn{6}{c}{ $\sigma(e^-e^+\to \mu^-\bar\nu_\mu u\bar d\,X)$(fb)}&
    \\\hline
    $\sqrt{s}\,[\mbox{GeV}]$& $\hat \sigma^{(3\slash 2)}$ & 
    $\Delta\hat \sigma^{{\rm C}\times[\rm{S+H}]}$ &  
    $\Delta\sigma^{\text{NLO-C}}$
    & $\Delta\sigma^{\rm{C}\times \text{decay}} $ 
    & $\Delta\sigma^{\rm{C}\times \text{res}} $ & 
      $\Delta\sigma^{\rm{C3}}$ &$\Delta\sigma^{\rm{C2}}$  \\\hline
    158 & $-0.001$ & $-0.116$ & 0.104& $-0.037$& 0.044  & 0.004 & 0.151 \\\hline
    161 &  0.147 & $-0.321$ & 0.226& $-0.091$& 0.324 & 0.010  & 0.437 \\\hline
    164 &  0.811 & $-0.417$ & 0.393&$-0.134$& 0.965 & 0.003  &0.399 \\\hline
    167 &  1.287 & $-0.389$ & 0.473 &$ -0.142$ &1.345 &0.001  &0.303 \\\hline 
    170 &  1.577 & $-0.354$ & 0.511&$-0.142$&1.561 &0.000& 0.246\\
      \hline
  \end{tabular}
  \end{center}
  \caption{Combined N$^{3\slash 2}$LO corrections (second column) and separate
    contributions from interference of single-Coulomb exchange with soft and
    hard corrections (third colum), renormalisation of the Coulomb
    potential (fourth column),
    interference of decay correction and single-Coulomb exchange 
    (fifth column), interference of residue correction and single-Coulomb 
    exchange (sixth column) and triple-Coulomb exchange 
    (all corrections are without ISR improvement).
    For comparison the NLO contribution from double-Coulomb exchange 
    (C2, second column)  are also shown.} 
  \label{tab:tabnum}
  \end{table}

  \begin{table}[p]
  \begin{center}
  \begin{tabular}{|c||c|c|c||c|c|}
    \hline&
    \multicolumn{4}{c}{ $\sigma(e^-e^+\to \mu^-\bar\nu_\mu u\bar d\,X)$(fb)}&
    \\\hline
    $\sqrt{s}\,[\mbox{GeV}]$ & Born &Born (ISR) & NLO &
    $\hat \sigma^{(3\slash 2)}$ & 
    $\sigma^{(3\slash 2)}_{\rm ISR}$ \\\hline
    158 &  61.67(2)& 45.64(2)&49.19(2)& $-0.001$ & 0.000   \\
            &  &  $[-26.0\%]$   & $[-20.2\%]$ &  [$-0.0$\permille] &  
    [+0.0\permille]    \\\hline
   161   & 154.19(6) &  108.60(4) &117.81(5)& 0.147 & 0.087 \\
       &   & $[-29.6\%]$ &$[-23.6\%]$ & [+1.0\permille] & [+0.6\permille] \\\hline
   164   & 303.0(1) & 219.7(1) &234.9(1) & 0.811  &0.544  \\
        & &   $[-27.5\%]$ & $[-22.5\%]$ &  [+2.7\permille] &[+1.8\permille]\\\hline
  167     & 408.8(2)&  310.2(1)&328.2(1)&  1.287& 0.936    \\
     &   & $[-24.1\%]$ &   $[-19.7\%]$&   [+3.1\permille] &[+2.3\permille] \\\hline
  170   & 481.7(2)& 378.4(2) &398.0(2)& 1.577  &1.207      \\
        & &    $[-21.4\%]$ &  $[-17.4\%]$ & [+3.3\permille]&[+2.5\permille] \\\hline
  \end{tabular}
  \end{center}
  \caption{
Two implementations of the N$^{3\slash 2}$LO corrections, which differ by the 
treatment of initial-state
radiation compared to the ``exact'' Born cross section without (second
column) and with (third column) ISR improvement and the NLO EFT result 
including ISR (fourth column). The relative
correction in brackets is given with respect to the Born cross
section in the second column. } 
  \label{tab:tabnum-1}
  \end{table}

\begin{figure}[t]
\begin{center}
\includegraphics[width=0.7\linewidth]{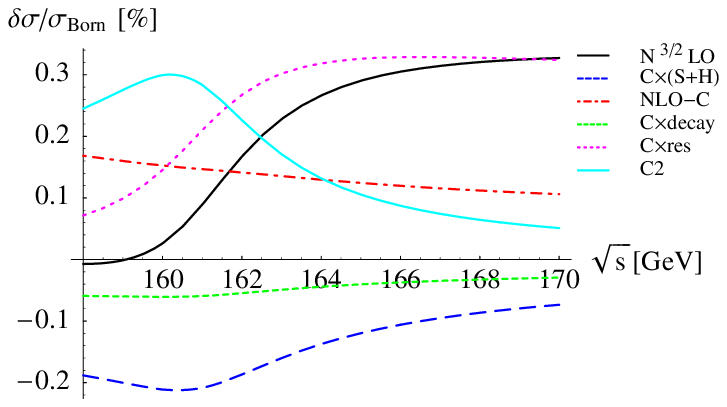}
\end{center}
\vspace*{-0.2cm}
\caption{Corrections relative to the Born cross section:
Combined N$^{3\slash 2}$LO (solid/black), interference of Coulomb with soft and hard corrections (long-dashed/blue), correction to the Coulomb potential (dash-dotted/red),  interference of Coulomb and decay corrections
 (short-dashed/green) and  interference of residue correction and single-Coulomb exchange (dotted/magenta). For comparison the NLO correction from double-Coulomb exchange (light solid/cyan) is also shown.}
\label{fig:results}
\end{figure}

In table~\ref{tab:tabnum} we show the numerical results for the individual
N$^{3\slash 2}$LO corrections to the unpolarised cross section, 
$\Delta \sigma^i=\Delta\sigma^i_{LR}/4$, at various
centre-of-mass energies near the $W$-pair production threshold.
In detail, the table contains corrections from interference of single-Coulomb 
exchange and soft/hard corrections
from~\eqref{eq:sigma-three-half} (third column),
from the NLO corrections to the
Coulomb-potential~\eqref{eq:bubble-gmu} (fourth column), from
interference of single-Coulomb exchange with decay and residue
correction in equations~\eqref{eq:delta-decay}
and~\eqref{eq:delta-residue} (fifth and sixth columns) and from the
triple-Coulomb correction in~\eqref{Coulomb3} (seventh column). 
The result~\eqref{eq:three-half-parton} for the sum of the different 
contributions is given in the second column. It is
also instructive to compare to the size of the pure two-Coulomb
exchange corrections given by the last term in~\eqref{CoulombCS}
(eighth column) that also arise from NNLO diagrams in the full
theory but appear already at NLO in the EFT power counting. 
Figure~\ref{fig:results} shows the size of
the individual corrections and the combined corrections relative to the
Born cross section in the full theory computed using a fixed-width
prescription.  It can be seen that individual corrections are 
comparable in magnitude to the second Coulomb correction but cancel to a
certain extent, in particular in the immediate threshold region where
the total N$^{3\slash 2}$LO correction is about one per-mille of the
Born cross section. Above threshold the residue correction dominates
and the total correction rises to about three per-mille of the Born
cross section.

The final results for the N$^{3\slash 2}$LO corrections including ISR
improvement (i.e. the result of inserting $\hat\sigma^{(3\slash 2)}$
into the convolution with the electron structure functions
in~\eqref{convolute}) is given in the last column of
table~\ref{tab:tabnum-1}, together with the previous result without
ISR improvement shown in the fifth column.  For comparison the Born
cross section in the fixed-width prescription with and without ISR
improvement generated using WHIZARD~\cite{Kilian:2007gr} and the
result of the NLO calculation of~\cite{Beneke:2007zg} (including ISR
improvement) are also shown.  The ISR improvement is seen to reduce
the N$^{3\slash 2}$LO corrections by about $40 \%$ at threshold and
$25 \%$ at $170\,\text{GeV}$.  Note that the effect is bigger than in
the case of the NLO corrections~\cite{Beneke:2007zg} that are reduced
by ISR improvement by about $20\%$ near threshold while there is
almost no effect at $170\,\text{GeV}$.  We remind the reader that 
when the N$^{3\slash 2}$LO terms presented here are added to the full 
1-loop calculation in the fixed-width or complex-mass 
scheme~\cite{Denner:2005es,Denner:2005fg} (as shown in 
appendix~\ref{converttoCMS}, the N$^{3/2}$LO result can be added to the
fixed-order NLO result in the complex-mass scheme without
modification), the double-Coulomb exchange terms C2 shown in 
table~\ref{tab:tabnum-1} must also be added, since these two-loop 
virtual effects, although part of the
NLO EFT calculation shown in table~\ref{tab:tabnum-1}, are not 
included in the fixed-order NLO 
calculation~\cite{Denner:2005es,Denner:2005fg}.

In~\cite{Beneke:2007zg} the impact of the interference of single-Coulomb 
exchange with hard and soft corrections on the $W$-mass
measurement was estimated as $[\delta M_W]\approx -5$ MeV. This
expectation was based on a naive estimate for these contributions in
eq.~(91) of~\cite{Beneke:2007zg} that corresponds to a correction to
the cross section of $\Delta\sigma^{3\slash
  2}_{\text{est.}}(161\text{GeV})\sim -0.27$fb. From
table~\ref{tab:tabnum} one sees that this correctly captures the order
of magnitude of the contribution from interference with soft
corrections and hard corrections to the production operator (third
column), that is, however, almost completely cancelled by the
correction to the Coulomb potential. Adopting the same procedure 
as in section~6.4 of~\cite{Beneke:2007zg} to estimate the 
shift of the $W$ mass, however assigning a relative error to each energy 
point that scales as one over the square root of the expected number
of events, we find that the impact of the
N$^{3\slash 2}$LO corrections on the $W$-mass measurement is 
about $3$ MeV (5$\,$MeV if the N$^{3\slash 2}$LO correction is not
convoluted with ISR), smaller than the targeted accuracy. Since other SM 
NNLO terms are expected to be even smaller, we may conclude that the 
(partonic) four-fermion cross section near the $W$-pair production
threshold is known with sufficient precision.

  \section{Towards the experimentally measured cross section}
\label{sec:cuts}
In measurements of the four-fermion production cross section,
cuts on momenta or angles of the observed particles have to
be applied in order to disentangle  signal and background
processes. While presently we do not aim at a general treatment of
such phase-space cuts using effective-theory methods, we would like to
estimate their effect on the corrections calculated in this article.

For orientation, we consider
the selection cuts used for the measurement of the four-fermion 
production cross section at $\sqrt s= 161$~GeV at 
LEP~\cite{Ackerstaff:1996nk,Abreu:1997sn,Acciarri:1997xc,Barate:1997mn}. 
To be definite, we use the cuts used by the L3 collaboration 
in~\cite{Acciarri:1997xc} to select the $q\bar q\mu\nu(\gamma)$ final state
that can be summarised as follows:
\begin{enumerate}[(i)]
\item\label{cut-mu} The muon momentum has to satisfy $|\vec p_\mu|>20$~GeV;
\item\label{cut-mass}
  the jet-jet invariant mass $M_{jj}$ and the invariant mass
  $M_{\mu\nu}$ of the muon-neutrino system have to satisfy
  $40 \text{ GeV}< M_{jj}< 120 \text{ GeV}$ and $M_{\mu\nu}>55$~GeV,
  respectively;
\item\label{cut-mu-j}
 the angle between muons and both hadronic 
jets must satisfy $\theta_{\mu j}>15$ degrees to suppress backgrounds from
$q\bar q (\gamma)$ production where muons arise as decay products of hadrons;
\item \label{cut-nu}
the polar angle of the missing-momentum vector has to satisfy
$|\cos\theta_\nu|<0.95$ to suppress $q\bar q (\gamma)$ events where the
missing energy arises from a photon lost in the beam pipe.
\end{enumerate}
To discuss the possible effects of these cuts
on the N$^{3\slash 2}$LO corrections, it has to be understood to which extent
they can be incorporated in the effective field theory framework. 
This will be discussed  first for the leading-order cross section
before the effect on the radiative corrections is considered.

We have studied the effect of the cuts on 
the Born cross section in the full SM  using WHIZARD~\cite{Kilian:2007gr}.
The numerical effects are shown in table~\ref{tab:cuts}.
These results do not include ISR improvement but we have checked that 
adding ISR does not change the relative impact of the cuts.
Although individual cuts are not very restrictive, their combined effect 
reduces the cross section by about 9 percent. 
The effect of the cut~(\ref{cut-mu}) 
is of the order of three per-mille and will not be considered further.
The remaining cuts fall into two categories: the cut~(\ref{cut-mass}) on the
invariant masses of pairs of decay products can be implemented in the 
effective-theory calculation as discussed in subsection~\ref{sec:mass-cut}. 
As shown there, these cuts do not affect the corrections calculated in this
article at all.
The  cuts ~(\ref{cut-mu-j}) and~(\ref{cut-nu})
are sensitive to angular distributions of the decay 
products of the $W$ bosons and are more problematic in the approach followed
in~\cite{Beneke:2007zg} and here. The uncertainty
on the theoretical prediction  introduced by these cuts 
is estimated in subsection~\ref{sec:angle-cuts}.

\begin{table}[t]
  \centering
  \begin{tabular}{|c|c|c|}
  \hline
  Cut & $\sigma_{\text{Born}} (e^-e^+\to \mu^-\overline{\nu}_\mu 
   u\overline{d})$(fb)& 
  $\sigma_{\text{cut}}/\sigma_{\text{tot}}$\\ \hline 
    -- & 154.18(5)& \\ \hline\hline
  $|\vec p_\mu|>20$~GeV& 153.71(5)& 99.69(5) $\%$   \\ \hline  
  $M_{\mu\nu}>55$~GeV, $40 \text{ GeV}< M_{jj}< 120 \text{ GeV}$
  & 150.61(5)& 97.68(5) $\%$
  \\ \hline
 $\theta_{\mu j}>15 \text{ degrees}$ &149.35(5)& 96.87(5) $\%$  \\ \hline
 $|\cos\theta_\nu|<0.95$ & 148.28(5) & 96.17(5) $\%$\\ \hline \hline
 all & 140.03(5) & 90.82(5) $\%$ \\ \hline
 \end{tabular}
 \caption{Effects of the phase-space cuts used
   in~\cite{Acciarri:1997xc} on the Born cross section in the full SM
   at $\sqrt s=161$ GeV without ISR improvement computed using WHIZARD. }
\label{tab:cuts}
\end{table}

\subsection{Invariant-mass cuts in the effective theory} 
\label{sec:mass-cut}

The precise treatment of invariant-mass cuts on the $W$-decay products
of the form
$-\Lambda^2_1<M_{f_if_j}^2-M_W^2<\Lambda^2_2$
in the effective theory depends on the scaling assigned to the ratio
$\Lambda/M_W$ with respect to the expansion parameter
$\delta \sim \Gamma_W/M_W$.  For the very loose cuts
in~(\ref{cut-mass}) above, it is appropriate to count $\Lambda/M_W \sim
1$ but we find it illuminating to consider also the possibility of
tighter cuts of the order 
$\Lambda/M_W\sim \sqrt{\Gamma_W/M_W}\sim \sqrt\delta$.\footnote{
In this respect it is interesting to mention that in the 
effective-theory treatment of top-pair production near
threshold~\cite{Hoang-lcws} the assumed hierarchy $\Gamma \ll \Lambda \ll M$
is closer to the second case.}  

To discuss the correct treatment of invariant mass cuts, 
recall that the total cross section is
extracted from appropriate unitarity cuts of the $e^-e^+$ forward-scattering 
amplitude~\cite{Beneke:2007zg}. The relevant diagrams
receive contributions from $W$-bosons with potential and hard loop
momenta. The potential region contributes starting at LO and is
reproduced in the effective theory by the matrix-element of
production/destruction operators as shown for leading order in~\eqref{LOamp},
 where the $W$-propagators are given by~\eqref{cor:delta}. 
As long as no soft-photon corrections are considered, the cut on the
invariant mass of the decay products
translates to cuts
on the momenta of the $W$s circulating in the loop of the form 
$-\Lambda_1^2<p_W^2-M_W^2<\Lambda^2_2$. These are implemented
by inserting a product of step-functions 
$\theta(\Lambda^2_2-p_W^2+M_W^2) \,\theta(\Lambda^2_1+p_W^2-M_W^2)$ in 
the cut loop integral.
We briefly comment on soft photons below.
The hard region contributes from N$^{1/2}$LO and is
reproduced in the effective theory by four-electron
operators~\cite{Beneke:2007zg}. The effect of invariant mass-cuts in
the hard region is included by taking the step functions 
into account in the calculation of the matching coefficients of the 
four-electron operators.
 We now discuss the two different possible scalings assigned to
 $\Lambda/M_W$ in turn. 

\paragraph{\it Loose cuts: $\Lambda\sim M_W$}
Consider the contribution of the potential region first.
Decomposing the $W$-loop momentum as $p_W=M_W v + r$ with $v=(1,\vec
0\,)$ and a 
residual potential momentum $r=(r^0,\vec r\,)\sim (\delta, \sqrt \delta)$, 
the step functions implementing the cuts are expanded as 
$\theta(\Lambda^2 \pm (2M_W r_0 -\vec r^2) )$. 
Since by assumption $\Lambda \gg r_0,\vec r^{\,2}\sim \delta$ the
momentum can be dropped in the step function so that
at leading order the cut can be neglected in the loop integrals
in the effective theory.
In the calculation of the matching coefficients of the four-electron
operators, on the other hand, the step functions
are operative since the $W$-loop momenta are hard and taken to 
be of the order $p_W^2\sim M_W^2\sim \Lambda^2$.
Therefore a loose cut with $\Lambda\sim M_W$ is taken into account entirely 
by modifying the matching coefficient of the four-electron operator, 
while loop integrals in the effective theory are to be performed without 
constraint on the $W$-momenta.

\paragraph{\it Tight cuts: $\Lambda\sim M_W \sqrt\delta$} Here the 
situation is reversed compared to the previous case: in the evaluation 
of cut loop integrals in the effective theory one has by assumption 
$|2 M_W r_0 -\vec r^{\,2}| \sim \Lambda^2$, so the step functions 
have to be taken 
into account. In the calculation of the matching coefficients of the 
four-electron operators, the $W$ momenta are taken to be hard and satisfy 
$|p_W^2 -M_W^2| \gg \Lambda^2$. Since $\theta(\Lambda^2-|p_W^2-M_W^2|)
\sim \theta(-|p_W^2-M_W^2|)=0$, tight cuts lead to 
vanishing matching coefficients. 
Therefore the invariant-mass cuts have to be taken into account in 
the loop calculations in the effective theory, while in the case 
of tight cuts the 
four-electron operators do not contribute to the cross section at all. 
Since $\Lambda^2 \sim M_W \Gamma_W$ some recombination prescription of 
soft photons and final state fermions has to be supplied.

\begin{figure}[t!]
\begin{center}
\includegraphics[width=0.7 \linewidth]{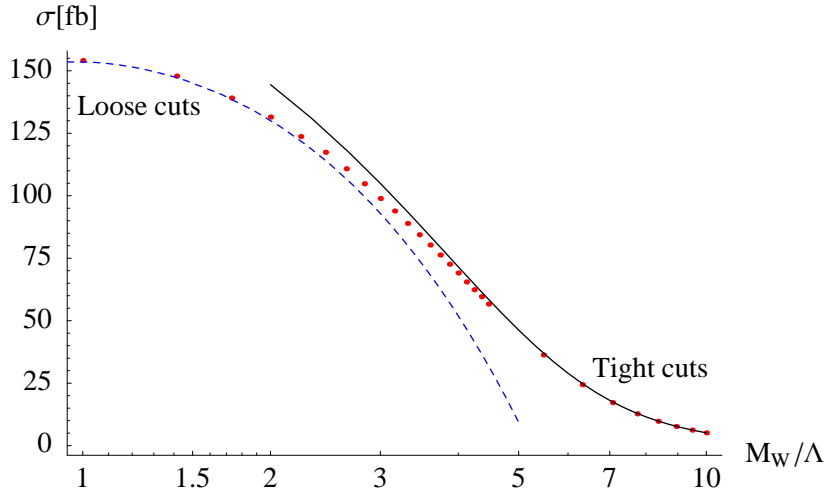}
\caption{Comparison of the Born cross section in the full SM computed 
with WHIZARD (red dots) with the effective-theory result for 
loose-cut implementation (dashed blue curve) and tight-cut implementation
(solid black curve). }
\label{fig:cut_cross}
\end{center}
\end{figure}

\paragraph{}
To verify that the above procedure is the appropriate treatment of 
invariant-mass cuts in the effective theory, we have computed the full
tree-level cross section for the process~\eqref{cor:proc} with
symmetric invariant-mass cuts $|M_{u\bar d}^2-M_W^2|<\Lambda^2$,
$|M_{\mu\bar \nu_\mu}^2-M_W^2|<\Lambda^2$ using WHIZARD, and compared
it to the effective theory cross section obtained using either the
counting $\Lambda\sim M_W$ or $\Lambda\sim M_W \sqrt{\delta}$.
The result shown in
figure \ref{fig:cut_cross} shows good agreement with the
effective-theory treatment in the regions where the respective
counting rule is appropriate. Implementing the cuts
of~\cite{Acciarri:1997xc} quoted in point (\ref{cut-mass}) at the
beginning of the section by modifying the matching coefficient of the
four-electron operator we obtain $\sigma_{\mbox{\tiny EFT}}(161
\mbox{GeV})=150.71 \, \mbox{fb}$ in good agreement with the WHIZARD
result of $\sigma(161 \mbox{GeV})=150.61 \pm 0.05 \, \mbox{fb} $.

We can now address the effect of  
cuts of the form  (\ref{cut-mass}) on radiative corrections.
The discussion above shows that at the order considered here the 
soft-photon and Coulomb-photon corrections computed in section~\ref{sec:eval} 
do not have to be modified by 
invariant mass cuts as long as they are of the order $\Lambda\sim M_W$.
The effects of such cuts on radiative corrections first appears 
in N$^{3\slash 2}$LO contributions to the four-fermion operators which we have not 
considered in this article since they are included in the NLO four-fermion 
calculation of~\cite{Denner:2005es,Denner:2005fg}. It is nevertheless
instructive to consider the expected effect using
the estimate of the correction to the total cross section~\cite{Beneke:2007zg} 
$\Delta\sigma_{4f}^{(3\slash 2)}(\sqrt s =161\text{GeV})\sim 
\alpha^ 4/(108 s_w^8 M^2_W)= 0.74$~fb, which is consistent with the difference 
of the EFT NLO calculation and~\cite{Denner:2005es,Denner:2005fg}.
Estimating the effect of the invariant-mass cuts on this contribution 
to be of the same order of magnitude of the one on the leading-order four-fermion operator, we have
\begin{equation}
\label{eq:three-half-cut}
\Delta \sigma_{4 f}^{(3\slash 2)} 
\left(1-\frac{\Delta \sigma_{4 f,\mbox{\tiny cut}}^{(1/2)}}{\Delta \sigma_{4 f}^{(1/2)}}\right)
\sim 0.03 \, \mbox{fb} \, .
\end{equation}   
Clearly the effect of the cut is totally negligible.

\subsection{Estimating the effect of angular cuts}
\label{sec:angle-cuts}

To discuss the treatment of cuts
 that involve angles between final state particles such
as~(\ref{cut-mu-j}) and~(\ref{cut-nu}) in the effective-theory framework
 we will first consider the effect of  a cut  on  the angle  between a
  final-state particle   and  the   beam   pipe  such   as~(\ref{cut-nu})
on the leading-order EFT description before we estimate the effect on the
corrections computed in the present paper.

To discuss the effect of angular cuts  we can use a variant of the effective
theory that includes the decay products~\cite{Beneke:2004xd}.
In   the
leading-order effective-theory description the $W$ bosons are produced
in  a $s$-wave  by  the operator~\eqref{cor:ver}.  At  this order  the
angular distribution of the $W$s is predicted to be isotropic. Due to
spin correlations the decay-product distributions are not isotropic, and
an  explicit  calculation  gives  (we consider  for  definiteness  the
neutrino angular distribution)
\begin{equation} \label{eq:neutrino_angle}
\frac{1}{\sigma^{(0)}_{\mbox{\tiny EFT}}} 
\frac{d \sigma^{(0)}_{\mbox{\tiny EFT}}}{d \cos \theta_{\nu}}
= \frac{3}{16} (1-\cos \theta_\nu) (3+\cos \theta_\nu) \, .  
\end{equation}
Computing the effect of the cut
(\ref{cut-nu}) using (\ref{eq:neutrino_angle}) gives a result
$\kappa = \sigma^{(0)}_{\mbox{\tiny EFT, cut}}/
\sigma^{(0)}_{\mbox{\tiny EFT}}=96.16 \, \%$
which is in very good agreement with the number given in table
\ref{tab:cuts}. 
As another example, consider the neutrino forward-backward asymmetry
$|(\sigma(\cos\theta_\nu>0)-\sigma(\cos\theta_\nu<0)|/\sigma_{\text{tot}}
=3/8$ computed from~\eqref{eq:neutrino_angle}
as compared to the result $\sim 30\%$ obtained using WHIZARD.

A closer look at the angular distributions generated by WHIZARD
reveals some deviations from the theoretical leading-order results,
for example a forward-backward asymmetry of the $W$-bosons
$|(\sigma(\cos\theta_W>0)-\sigma(\cos\theta_W<0)|/\sigma_{\text{tot}}\sim
30 \%$. These deviations are mainly to be attributed to non-resonant
momentum configurations that are incorporated in the four-fermion
operators that are formally of order N$^{1/2}$LO. At threshold the
effect of the four-fermion operators is about 40 percent of the
leading order EFT contributions~\cite{Beneke:2007zg}, which is
consistent with the order of magnitude of the observed asymmetries.
The WHIZARD results show an isotropic distribution of the $W$s if at
the same time invariant mass-cuts in the `tight-cut' regime are
applied where the four-electron operators vanish, as shown in
section~\ref{sec:mass-cut}

Turning to the corrections computed in this paper, we first note that
the Coulomb correction neither modifies the angular distribution nor
the polarization of the $W$ bosons. The same is true for the one-loop
corrections to the production operator $C^{(1)}_{p,LR}$ and the
hard-collinear corrections considered in section~\ref{sec:hard} that
can be both factorized from the Coulomb Green function.  Therefore
for all corrections considered in this paper apart from the soft
corrections computed in section~\ref{sec:soft} the effect of the
angular cuts will be the same as for the leading-order result. Since
the numerical effect of the N$^{3\slash 2}$LO corrections on the total
cross section is in the one to three  per-mille range, the effect of the
angular cuts (which reduce the cross section by about 7 percent) on
these corrections will be completely negligible.

We expect that this conclusion is not significantly changed by
the soft corrections. First, the components of the momentum of the 
soft emitted photon are $O(\Gamma_W)$ (see~\eqref{scale}), whereas the
momenta of the incoming and 
outgoing  fermions are collinear with an energy of $O(M_W)$.
Also the soft photons couple to collinear fermions with eikonal
vertices $\sim k^\mu$, that do not change the polarization
of the particle the photon is attached to.
Even if these arguments would underestimate the effect, the numerical 
results in the fourth column of table~\ref{tab:tabnum} 
show that a 10 per-cent uncertainty on the effect of the cuts on the 
soft corrections translates to an uncertainty of the cross section
much below $0.1$~fb and is therefore not relevant for the desired 
accuracy needed for the mass measurement. In summary, while a fully 
differential calculation of the N$^{3\slash 2}$LO corrections with 
the EFT method is not currently possible, the mild cuts relevant to 
the $W$ mass measurements together with the smallness of the 
correction allow us to conclude that adding 
the inclusive N$^{3\slash 2}$LO cross section estimate to the 
NLO result is sufficient in practice. 

\section{Conclusions}
\label{sec:conc}

The aim of this paper has been to ascertain that the four-fermion
production cross section near the $W$-pair production threshold can be
calculated with an accuracy that matches the estimated $6\,$MeV 
error in the $W$-mass measurement from 
an energy scan in $e^- e^+$ collisions, which requires 
the consideration of NNLO terms in the Standard Model. 
The effective field theory approach to unstable-particle production 
makes it possible to investigate the relevant part of the Standard
Model NNLO diagrams without facing the difficulties of a full
calculation, which would be overwhelming at present, and to identify 
a set of parametrically enhanced NNLO corrections, all associated 
with the strong electromagnetic Coulomb attraction of the intermediate
$W$ bosons. We obtain compact analytic expressions for the total 
cross section, and a result for soft, hard and collinear 
radiative corrections that extends to the presence of any number of 
Coulomb photon exchanges. The methods used in this calculation 
may be useful in other processes involving unstable particles 
such as top quarks or new heavy states. 

We find that individual correction terms to the four-fermion cross
section are in the $0.3\%$ range, leading
to shifts of the $W$ mass around $5\,$MeV. However, the combined
effect is somewhat smaller. We may thus conclude that the inclusive 
partonic four-fermion cross section near the $W$-pair production
threshold is known with sufficient precision. We then 
investigated the impact of cuts, showing how to systematically 
include invariant-mass cuts into the EFT in two cases differing by 
the scaling of the cut with the other parameters of the problem. 
The cuts applied in the measurement of the $W$-pair production cross
section at LEP2 loose only about 10\% of the total cross section. 
In view of the smallness of the corrections found here, no detailed 
differential cross section calculations beyond NLO appear to be 
necessary. Our result should then be combined with NLO calculations of 
the four-fermion cross section in the SM in the complex mass-scheme 
\cite{Denner:2005es,Denner:2005fg} as described in  
appendix~\ref{converttoCMS}. 

The partonic four-fermion cross section must be convoluted with
electron structure functions that sum large collinear logarithms. 
As discussed in~\cite{Beneke:2007zg} to achieve the required precision
for the $W$-mass measurement it will be necessary to improve 
the treatment of initial-state radiation that was applied at LEP to 
account consistently for next-to-leading logarithms. This problem is
common to all high-precision studies in high-energy $e^- e^+$ 
collisions.

\section*{Acknowledgement}
This work is supported by the DFG Sonder\-forschungsbereich/Transregio~9 
``Computergest\"utzte Theoretische Teilchenphysik''. 
Feynman diagrams have been drawn with the packages 
{\sc Axodraw}~\cite{Vermaseren:1994je} and 
{\sc Jaxo\-draw}~\cite{Binosi:2003yf}.
\appendix
\section{Renormalisation of the Coulomb potential by hard corrections}
\label{sec:hard-coulomb}
In this appendix we discuss several technical aspects related to the
hard corrections to the Coulomb potential used in section~\ref{sec:bubble} and
their dependence on the renormalisation scheme.  The calculation of
the corrections to the single-Coulomb exchange requires a matching
calculation where one computes the renormalised
$W^+W^-\to W^+W^-$ NLO scattering amplitude in the full theory for
$(p_1+p_2)^2=4M_W^2$ and compares to the one-loop $\Omega^+\Omega^-\to
\Omega^+\Omega^-$ amplitude in the effective theory. 
Equivalently, 
the full-theory calculation can be split into contributions from different
momentum regions. The relevant regions are the hard, potential, soft
and semi-soft regions.  The only contribution that will not be
reproduced by diagrams in the EFT is that from the hard region, so for
the matching calculation it is sufficient to calculate the hard
corrections. At leading order in the non-relativistic expansion 
only the corrections to the single-Coulomb exchange diagram contribute.
In~\ref{sec:counter-coulomb} we define our renormalisation
conventions, the hard corrections to the process $W^+W^-\to W^+ W^-$
are discussed in~\ref{sec:hard-ww}. The relevant results in the
$\alpha(M_Z)$ and $G_\mu$ input parameter schemes are collected
in~\ref{sec:convert}.

\subsection{Charge renormalisation}
\label{sec:counter-coulomb}

The lowest perturbative scale relevant near the $W$-pair production
threshold is the $W$ width $\Gamma_W$, so we will employ a 
renormalisation scheme $S$ for the electric charge that is not sensitive to 
smaller scales, in particular not to the
light-fermion masses.
In practice we will use $\alpha(M_Z)$ or the Fermi constant $G_\mu$ as
input parameter (see e.g. \cite{Dittmaier:2001ay}), but for the moment we will leave the renormalisation scheme
unspecified.  
  Following the
renormalisation conventions of~\cite{Beneke:2007zg} the overall
one-loop counterterm of the $W^+W^-\to W^+W^-$ amplitude in the full
theory in a given charge-renormalisation scheme $S$ is given by
\begin{equation}
\label{eq:ctterm}
\Delta^S_{\text{counter}}={\rm [tree]} \times \left( 2\, \delta Z^S_e +
2 \delta Z_{W} \right) ,
\end{equation}
since the tree-level single photon exchange diagram (denoted by [tree])
is proportional to 
$e^2=(4\pi\alpha)^2$ and has four external $W$-legs. 

In the following we write the charge counterterm in a given scheme $S$
 as the counterterm in the 
$\alpha(0)$ scheme and a finite scheme-dependent shift
\begin{equation}
   \delta Z_e^S= \delta Z_e^{\alpha(0)}
-\frac{1}{2}\Delta \alpha^{S},
\end{equation}
with the explicit expressions for the charge counterterm in the $\alpha(0)$ scheme~\cite{Denner:1991kt}\footnote{
Here all conventions are the same as 
in~\cite{Denner:1991kt} apart from replacing $\Sigma\to -\Pi$. 
Note that~\cite{Denner:1991kt} defines the
vacuum polarisation $\Pi_{AA}(k^2)=\Sigma_{AA}(k^2)/k^2$ which we
don't use in the following.}
\begin{equation}
   \delta Z_e^{\alpha(0)}
=-\frac{1}{2}\delta Z_{AA}-\frac{s_w}{c_w}\frac{1}{2}\delta Z_{ZA}
= -\frac{1}{2}\frac{\partial\Pi_T^{AA}(k^2)}{\partial k^2}|_{k^2=0} 
+\frac{s_w}{c_w}\frac{\Pi_T^{AZ}(0)}{M_Z^2},
\end{equation}
where the transverse self-energies are defined by the decomposition
\begin{equation}
  \Pi^{VV}_{\mu\nu}(q)=
  \left(g_{\mu\nu}-\frac{q_\mu q_\nu}{q^2}\right)\Pi^{VV}_T(q^2)+
  \frac{q_\mu q_\nu}{q^2}\Pi^{VV}_L(q^2).
\end{equation}
For the argument given below we assume that light-fermion masses are
used as IR regulators in the $\alpha(0)$ scheme, but this will only be
used in intermediate steps and the dependence on the light-fermion
masses drops out in the end.

\subsection{Hard corrections}
\label{sec:hard-ww}
\begin{figure}[ht]
  \begin{center}
  \includegraphics[width=0.7\textwidth]{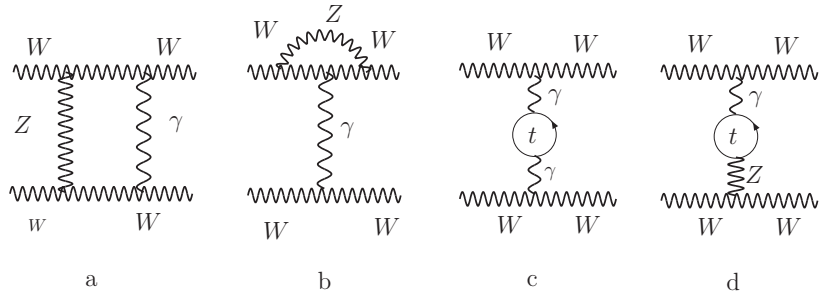}
  \caption{Sample diagrams in the full SM contributing to the hard corrections to the $WW\to WW$ subprocess}
  \label{fig:hard-coulomb}
  \end{center}
  \end{figure}

As discussed at the beginning of this appendix we need to calculate 
the hard corrections to the  $W^+W^-\to W^+ W^-$ amplitude for external 
momenta directly at threshold. 
The contributing diagrams are of the form of box corrections
(figure~\ref{fig:hard-coulomb}a), vertex corrections
(figure~\ref{fig:hard-coulomb}b) and self-energy insertions
(figure~\ref{fig:hard-coulomb}c,d).
The box corrections with a hard
loop momentum do not contribute at the order we are considering since near
threshold they
are suppressed by a factor $v$ compared to the diagrams with a  Coulomb photon.
This leaves the vertex and the bubble corrections.

  Expanding the vertex-correction diagrams of the form of
  figure~\ref{fig:hard-coulomb}b in the region where the photon and
  the $W$s attached to the vertex sub-loop are potential and the
  momentum running in the vertex loop is hard, one obtains the
  single-Coulomb exchange diagram with an insertion of the one-loop $WW\gamma$
  vertex function evaluated with on-shell external momenta.  The
  renormalised on-shell vertex function vanishes in the conventional
  on-shell renormalisation scheme where $\alpha(0)$ is used as input
  parameter.\footnote{This relies on `charge universality' in the
    standard model, i.e. on the fact that the on-shell electron-photon
    vertex and the on-shell $W$-photon vertex receive the same
    radiative corrections~\cite{Bohm:2001yx}.}  Therefore the
  diagrams with insertion of a one-loop \emph{unrenormalised}
  $WW\gamma$ vertex are given by the negative of the corresponding
  counterterm in the on-shell renormalisation scheme:
\begin{equation}
\Delta_{\text{vertex}}=
  [\text{tree}]\times (-2\delta^{os}_{WW\gamma})= [\text{tree}]\times
(-2) \left(\delta Z^{\alpha(0)}_e +\delta Z_W +\frac{1}{2}\delta Z_{AA}-
\frac{1}{2}\frac{c_w}{s_w}\delta Z_{ZA}\right). 
\end{equation}

A similar argument shows that in the hard region the self-energies in
figures~\ref{fig:hard-coulomb}c/d are evaluated at zero external
momentum.  Since the (unrenormalised) one-loop photon self-energy
vanishes at zero external momentum the first non-vanishing contributions of
diagrams of the form~\ref{fig:hard-coulomb}c comes from expanding 
the self-energy to first order:
\begin{equation}
\label{eq:bubble-hard}
  \Delta_\gamma=[\text{tree}] \times 
(i\Pi^{AA,\text{hard}}_{T}(k^2))\frac{-i}{k^2}
  =[\text{tree}] \times
 \left(\frac{\partial \Pi^{AA}_{\text{heavy}}(k^2)}{\partial k^2}|_{k^2=0} \right),
\end{equation}
where $\Pi^{AA}_{\text{heavy}}$ includes all particles except the light
fermions.  Here it was used that for a hard loop momentum the light-fermion 
masses must be set to zero so the loop integral is
scaleless and only the heavy particles contribute to the hard part of the 
self-energy at zero external momentum. The unrenormalised
photon-$Z$ mixing diagrams give (suppressing the transverse projectors)
\begin{eqnarray}
\Delta_{\gamma/Z}
&=& \lbrack\text{tree}\rbrack \times 2 \,(i \Pi^{AZ}_T(0))
\,\frac{-i}{k^2-M_Z^2}\;
\frac{g_{ZWW}}{g_{\gamma WW}}
\nonumber\\
&\Rightarrow& 
[\text{tree}] \times \left(-2\frac{\Pi_{T,\text{heavy}}^{AZ}(0)}{M_Z^2} \right)
\left(-\frac{c_w}{s_w}\right)
=  [\text{tree}] \times (\delta Z_{ZA})\left(-\frac{c_w}{s_w}\right),
\end{eqnarray}
since for a potential momentum $k^2\ll M_Z^2$. Here the on-shell scheme 
definition 
$\delta Z_{ZA}= -2 \Pi_T^{AZ}(0)/M_Z^2$ has been used.
We also used that massless fermion loops do not contribute 
to the $\gamma-Z$ mixing at zero external momenta, as can be seen 
from the explicit one-loop result in~\cite{Denner:1991kt}. 
Therefore $\Pi_{T,\text{heavy}}^{AZ}(0)=\Pi_{T}^{AZ}(0)$.

Adding the vertex correction, the $\gamma/Z$ mixing and the 
counterterm~\eqref{eq:ctterm} (where `tree' is again only the photon 
exchange diagram) 
one obtains the hard correction in the renormalisation scheme $S$:
\begin{equation}
 \Delta^{S}_{\text{hard}}=
\Delta_{\text{vertex}}+\Delta_{\gamma}
+\Delta_{\gamma/Z}+\Delta_{\text{counter}}^S
\equiv [\text{tree}]\times \delta_{\text{hard}}^S , 
\end{equation}
where we defined the correction factor
\begin{equation}
\label{eq:hard-coulomb}
\delta_{\text{hard}}^S= 2(\delta Z^S_e-\delta Z_e^{\alpha(0)})
-\delta Z_{AA}+\frac{\partial \Pi^{AA}_{\text{heavy}}(k^2)}{\partial k^2}|_{k^2=0} 
= -\Delta \alpha^{S}-
\frac{\partial \Pi^{AA}_{f\neq t}(k^2)}{\partial k^2}|_{k^2=0} . 
\end{equation}
In this expression the derivative of the light-fermion 
contribution to the photon self-energy
 depends on light-fermion masses used as regulators in the $\alpha(0)$ scheme. 
This dependence will be cancelled by a similar term in the conversion factor
$\Delta\alpha^S$ for a scheme $S$ that is not sensitive to scales below $\Gamma_W$.

Alternatively, the  result~\eqref{eq:hard-coulomb} is obtained by considering the individual renormalised contributions instead of
applying the overall counterterm~\eqref{eq:ctterm}. 
In this case 
\begin{equation}
\Delta^S_{\text{hard}}=\Delta_{\text{vertex}}^r+\Delta_{\gamma}^{r}+
\Delta_{\gamma/Z}^{r}.
\end{equation}
Since the on-shell 
renormalised $Z-\gamma$ mixing two-point function vanishes at zero momentum 
we have $\Delta^{r}_{\gamma/Z}=0$. The renormalised one-loop correction to 
the $WW\gamma$ vertex vanishes in the on-shell scheme so the only non-vanishing contribution comes from the change in the charge-counterterm:
\begin{equation}
  \Delta^r_{\text{vertex}}=2\,[\text{tree}]\times 
  (\delta Z^S_e-\delta Z_e^{\alpha(0)}).
\end{equation}
The renormalised self-energy correction to the photon-exchange is given by~\eqref{eq:bubble-hard} and the corresponding counterterm:
\begin{equation}
  \Delta^r_\gamma
= [\text{tree}]\times \left(\frac{\partial \Pi^{AA}_{\text{heavy}}(k^2)}{\partial k^2}|_{k^2=0}
-\delta Z^{AA}\right).
\end{equation}

\subsection{Formulas for the $\alpha(M_Z)$ and $G_\mu$ schemes}
\label{sec:convert}
We now specialise the result~\eqref{eq:hard-coulomb} to the two schemes used
in the main text.
In the $\alpha(M_Z)$ scheme the finite shift of the charge counterterm
is given by (see e.g. \cite{Denner:1991kt})
\begin{equation}
\Delta\alpha^{M_Z}= 
-\frac{\partial\Pi^{AA}_{f\neq t}(k^2)}{\partial k^2}|_{k^2=0} +
\frac{\re\,\Pi_{f\neq t}^{AA}(M_Z^2)}{M_Z^2}.
\end{equation}
Inserting this definition  into~\eqref{eq:hard-coulomb}, 
the sensitivity on the light-fermion masses drops out and one 
obtains the final result for the hard corrections to the Coulomb
potential (charge counterterm) in 
the $\alpha(M_Z)$ scheme:
\begin{equation}
\label{eq:delta-mz}  
  \delta_{\text{hard}}^{\alpha(M_Z)}= 
  -\frac{\re\,\Pi_{f\neq t}^{AA}(M_Z^2)}{M_Z^2}.
\end{equation}

In the $G_\mu$  scheme the shift in the counterterm is instead given by
the correction to muon decay, $\Delta r$,
\begin{align}
\Delta \alpha^{G_\mu}=\Delta r&=
  -\frac{\partial\Pi^{AA}_T(k^2)}{\partial k^2}|_{k^2=0} 
- 2\frac{\delta s_w}{s_w}-
2\frac{c_w}{s_w}\frac{\Pi_T^{AZ}(0)}{M_Z^2} - 
\frac{\Pi^{W}_{T}(0)- {\rm Re}\, \Pi^{W}_{T}(M_W^2)}{M_W^2} +\delta r, \\
\frac{\delta s_w}{s_w}&=\frac{1}{2}\frac{c_w^2}{s_w^2}
\left(\frac{\re \Pi_T^W(M_W^2)}{M_W^2}-\frac{\re\Pi_T^Z(M_Z^2)}{M_Z^2}\right),
\\
\delta r &= \frac{\alpha}{4\pi s_w^2}
    \left(6+\frac{7-4 s_w^2}{2 s_w^2} \ln c_w^2\right).
\end{align}
The explicit result for the hard corrections in the $\alpha_{G_\mu}$ scheme 
from~\eqref{eq:hard-coulomb} reads
\begin{equation}
\label{eq:delta-gmu}  
\begin{aligned}
  \delta_{\text{hard}}^{G_\mu}&= -\Delta r -
\frac{\partial \Pi^{AA}_{f\neq t}(k^2)}{\partial k^2}|_{k^2=0} \\
&=\frac{\partial\Pi^{AA}_{\text{heavy}}(k^2)}{\partial k^2}|_{k^2=0} 
+ 2\frac{\delta s_w}{s_w}
+2\frac{c_w}{s_w}\frac{\Pi_T^{AZ}(0)}{M_Z^2} +
\frac{\Pi^{W}_{T}(0)- {\rm Re}\, \Pi^{W}_{T}(M_W^2)}{M_W^2} -\delta r .
\end{aligned}
\end{equation}
To convert the result from the $\alpha(M_z)$ scheme to the 
$G_\mu$ scheme we have to add
\begin{eqnarray}
\delta_{\alpha(M_Z)\to G_\mu} &=&
 \delta^{G_\mu}_{\text{hard}}-\delta_{\text{hard}}^{\alpha(M_Z)}
=  \frac{{\rm Re}\,\Pi_{f\neq t}^{AA}(M_Z^2)}{M_Z^2}+
\frac{\partial\Pi^{AA}_{\text{heavy}}(k^2)}{\partial k^2}|_{k^2=0} 
\nonumber\\
&&\hspace*{0cm}
+ \,2\frac{\delta s_w}{s_w}
+2\frac{c_w}{s_w}\frac{\Pi_T^{AZ}(0)}{M_Z^2} +
\frac{\Pi^{W}_{T}(0)- {\rm Re}\, \Pi^{W}_{T}(M_W^2)}{M_W^2} -\delta r .
\label{eq:delta-mz-gmu}
\end{eqnarray}
Explicit expressions for the self-energies appearing in these quantities
 can be found for instance in~\cite{Denner:1991kt}.

\section{Conversion to the complex-mass scheme}
\label{converttoCMS}

In the main text we describe the calculation of those N${}^{3/2}$LO
correction in the EFT framework, which are not already included in a 
NLO calculation in the full Standard Model. The implicit assumption is
that the NLO diagram corresponding to exchange of a single (Coulomb)
photon between the $W$ bosons is calculated in the on-shell scheme and
defined with a fixed-width
prescription for the $W$ propagators.\footnote{Since all 
N${}^{3/2}$LO corrections calculated in the paper are $O(\alpha)$ 
corrections to the NLO diagram with a single-Coulomb photon, other NLO
diagrams are not relevant.} Here we discuss the changes that should be
made, if this diagram is calculated as part of a complete NLO SM
calculation in the complex-mass scheme defined as 
in~\cite{Denner:2005es,Denner:2005fg}.

Since the complex-mass scheme uses the same $W$-propagators as the
fixed-width scheme, the differences arise only from the different
renormalisation conventions for $\alpha_{ew}$ and $\alpha$, and the 
different treatment of $\Pi^{(1,1)}$ from $W$-field renormalisation. 
The different coupling definitions lead to a change in the value of 
$c_{p,LR}^{(1,{\rm fin})}$ in the $C\times H$ correction
(\ref{prodver}) and an
additional term involving $\delta_{G_\mu\to \alpha_{\rm CMS}}$ in 
the NLO Coulomb potential term (\ref{eq:bubble-gmu}). 
These compensate exactly for the
difference between the coupling prefactors 
$\alpha_{ew}^2\alpha$ in the single-Coulomb exchange diagrams 
in the complex-mass scheme and the $G_\mu$ scheme. Thus when the NLO 
SM calculation is done in the complex-mass scheme, we must add 
\begin{equation}
\Delta \sigma^{C1}_{LR}-[\Delta \sigma^{C1}_{LR}]_{\rm CMS}
\label{convertcms}
\end{equation}
expanded to order $\alpha_{ew}^2\alpha^2$ to the 
N$^{3/2}$LO calculation performed in the main text. 
However, at the indicated order the 
correction~\eqref{convertcms} actually vanishes. To see this, 
extract the relevant coupling constants from the LO scattering 
amplitude for four-fermion production $\mathcal{A}^{(0)} \equiv \alpha_{ew} 
\tilde{\mathcal{A}}^{(0)}$, and the amplitude including 
single-Coulomb exchange,  $\mathcal{A}^{(1/2),C1} \equiv \alpha_{ew}\alpha 
\tilde{\mathcal{A}}^{(1/2),C1}$. 
The correction to the cross section then involves the quantity
\begin{equation}
2 \,\re\left[{\mathcal{A}}^{(0)}\left(\mathcal{A}^{(1/2),C1}\right)^*\right]
= 2 \,|\alpha_{ew}|^2 \,\re\left[\alpha^* \tilde{\mathcal{A}}^{(0)}
 \left(\tilde{\mathcal{A}}^{(1/2),C1}\right)^*\right].
\end{equation}
In the implementation of the complex-mass scheme of~\cite{Denner:2005fg} the imaginary part of the electromagnetic charge arises only through 
complex masses entering loop integrals in the charge counterterm. 
Therefore the imaginary part of $\alpha$ 
involves terms of the form
 $\alpha^2 \,\im[\,\log((M_W^2- i M_W\Gamma_W)/\mu^2)]
\sim \alpha^2 \,\Gamma_W/M_W \sim \alpha^2 \alpha_{ew}$
and does not contribute at order $\alpha_{ew}^2\alpha^2$ 
in~\eqref{convertcms} due to the extra factor of 
$\Gamma_W/M_W\sim \alpha_{ew}$. Similarly the difference between 
$\alpha_{ew}^2$ and $|\alpha_{ew}|^2$ is of higher order, 
$\alpha_{ew}^2 (\Gamma_W/M_W)^2 \sim \alpha_{ew}^4$.

A change in the renormalisation convention that changes the
renormalised value of $\Pi^{(1,1)}$ affects the calculation of the 
residue correction, but also the value of $c_{p,LR}^{(1,{\rm fin})}$ and 
the NRQED $WW\gamma$ vertex, 
which depend on the $W$-field renormalisation convention. The two
exactly compensate each other as can be seen as follows: a change in
the real part of $\Pi^{(1,1)}$ changes the residue correction by an
amount that is directly proportional to $\Delta \sigma^{C1}_{LR}$, 
as is the term from the change of $c_{p,LR}^{(1,{\rm fin})}$, so the 
cancellation is obvious. A change in the imaginary part of
$\Pi^{(1,1)}$ is more subtle, since it affects the imaginary part of 
$c_{p,LR}^{(1,{\rm fin})}$. We dropped this imaginary part in the
on-shell scheme, since it does not correspond to cuts involving the
four-fermion final state, but this is no longer correct if the field
renormalisation is complex. In this case one must include the
imaginary part of $c_{p,LR}^{(1,{\rm fin})}$ that comes from
$\Pi^{(1,1)}$, in which case the $C\times H$ correction is no longer 
proportional to $\Delta \sigma^{C1}_{LR}$. Rather, as in the
calculation of the residue correction, one obtains the integral 
(\ref{rescorr}). We thus see again that the change in  $\Pi^{(1,1)}$ is
compensated by the one in $c_{p,LR}^{(1,{\rm fin})}$. 
A similar argument applies to the change in the 
$WW\gamma$ vertex. Hence no further
correction is needed to convert our
result to the case when the full NLO calculation is done in the
complex-mass scheme.


\end{document}